\documentclass[aps,pra,twocolumn,longbibliography,superscriptaddress,10pt]{revtex4-2}

\usepackage{amsmath,amssymb,bm}
\usepackage{graphicx}
\usepackage{dcolumn}
\usepackage{hyperref}
\usepackage[capitalise]{cleveref}
\makeatletter
\providecommand\@fourthoffive[5]{#4}
\providecommand\@fifthoffive[5]{#5}
\makeatother

\newcommand{\half}{\tfrac{1}{2}}
\newcommand{\Lz}{\hat{L}_z}
\newcommand{\avg}[1]{\langle #1 \rangle}
\newcommand{\Veff}{V^{\mathrm{eff}}}
\newcommand{\Rbox}{R_{\mathrm{box}}}
\newcommand{\aho}{a_{\perp}}

\newcommand{\etaGP}{\eta_{\mathrm{GP}}}
\newcommand{\Evar}{E_{\mathrm{var}}}
\newcommand{\lvar}{\ell_{\mathrm{var}}}
\newcommand{\eps}{\varepsilon}
\newcommand{\Mcore}{M_{\mathrm{core}}}
\newcommand{\Ncore}{N_{\mathrm{core}}}
\newcommand{\cE}{\mathcal{E}}
\newcommand{\R}{\mathbb{R}}
\newcommand{\dd}{\mathrm{d}}
\newcommand{\x}{\mathbf{x}}
\newcommand{\y}{\mathbf{y}}
\newcommand{\uu}{\mathbf{u}}
\newcommand{\vv}{\mathbf{v}}
\newcommand{\dvar}{\delta_c^{\mathrm{var}}}
\newcommand{\dGP}{\delta_c}
\newcommand{\epsgrad}{\gamma_{\nabla}}
\newcommand{\gflow}{\gamma_{v}}
\newcommand{\Lsq}{L_{\mathrm{sq}}}
\newcommand{\tcell}[2]{\parbox[t]{#1}{\raggedright #2\strut}}

\begin{document}

\title{Rotating two-dimensional Bose--Einstein condensates in power-law traps:\\
collapse, Thomas--Fermi scaling, and the inverse-square endpoint}

\author{Bao-Duy Le}
\email{24130012@student.hcmus.edu.vn}
\affiliation{Faculty of Physics and Engineering Physics, University of Science, Ho Chi Minh City, Vietnam}
\affiliation{Vietnam National University, Ho Chi Minh City, Vietnam}
\author{Dinh-Thi Nguyen}
\email{ndthi@hcmus.edu.vn}
\affiliation{Faculty of Mathematics and Computer Science,
University of Science, Ho Chi Minh City, Vietnam}
\affiliation{Vietnam National University, Ho Chi Minh City, Vietnam}

\date{\today}

\begin{abstract}
In two dimensions, the kinetic and contact energies scale identically under dilation,
a trap $V=r^s/s$ as $\lambda^{-s}$. We follow this balance through
the power-law family $-2\le s\le4$, $s\neq0$, for both signs of the coupling $g$, with
rotation, using scaling bounds, radial solutions, Bogoliubov spectra,
and two-dimensional imaginary-time states. Attraction collapses the cloud at the Townes
coupling $g_{*}$ in every trap, with width $\ell\propto\delta^{1/(s+2)}$,
$\delta=1-|g|/g_{*}$; repulsion spreads it to a Thomas--Fermi (TF) radius
$\propto g^{1/(s+2)}$. A rotating collapse leaves the axis once the rotation frequency
$\Omega$ exceeds its dipole frequency, which softens only for $s>2$; in a quartic trap
this happens at $\delta\simeq0.32\,\Omega^6$, within $3\%$ of the Bogoliubov threshold;
it then collapses on the minimum ring of the rotating-frame potential, which also
empties a repulsive TF cloud at $\Omega=(3g/\pi)^{1/6}$. For
repulsive vortex states the TF energy error splits exactly into a density-gradient and
a flow part of comparable or larger size. At $s=-2$ only the cutoff sets a length: the
energy is exactly covariant with it, the condensate falls to the center for both signs,
and repulsion leaves a halo of density $\propto1/(gr^2)$ whose extent grows exponentially
with $g$. Near collapse, split-step imaginary time has a first-order stationary bias
invisible to residual tests.
\end{abstract}

\maketitle

\section{Introduction}
\label{sec:intro}

Optical potentials now give access to trapped Bose--Einstein condensates (BECs)
well outside the harmonic paradigm: uniform boxes~\cite{Navon2021,Gauthier2016},
quartic-modified traps that admit rotation beyond the centrifugal
limit~\cite{Fetter2001,KavoulakisBaym2003,Bretin2004}, and engineered attractive dimples or
atom--surface potentials~\cite{Antezza2004}. Rotation fills such clouds with vortex
lattices~\cite{Madison2000,AboShaeer2001,Fetter2009}, and Feshbach tuning reaches the
attractive side up to collapse~\cite{Sackett1998,Donley2001}. The power-law family
$V_s(r)=r^s/s$ interpolates across these geometries with one exponent; its singular side
$s<0$ is an attractive well whose inverse-square member $s=-2$ is the marginal potential
of quantum mechanics~\cite{Case1950,LandauLifshitzQM,Camblong2001}.

One scaling argument organizes the whole family. Under the norm-preserving dilation
$\psi(\x)\to\lambda\psi(\lambda\x)$ of a two-dimensional order parameter, the kinetic and
contact energies both scale as $\lambda^2$~\cite{PitaevskiiRosch1997}, the trap energy as
$\lambda^{-s}$, and the angular momentum $\avg{\Lz}$ is invariant. The size $L$ of the
cloud is therefore set by balancing the trap against the scale-invariant pair,
$L^{s+2}\propto K$, where $K$ is the net coefficient of the two $\lambda^2$ terms. For
repulsion $K$ is dominated by the contact energy and grows with $g$: the cloud spreads to
the Thomas--Fermi (TF) radius $\propto g^{1/(s+2)}$, the trap acts through its shape over
the whole cloud, and the kinetic energy is a correction. For attraction $K$ is the small
difference of the two terms, and three consequences follow. (i)~Whether the condensate
collapses is decided by the contact coupling alone: the threshold is the Townes value
$g_{*}$~\cite{Townes1964,Weinstein1983} for every trap and for every rotation speed at
which the trap still confines.
(ii)~How it approaches the threshold is decided by the trap, and only through its local
shape where the cloud concentrates. (iii)~At $s=-2$ the trap scales like the pair itself,
so no balance and no length are left for either sign of $g$; only a cutoff can supply one.
We follow these consequences through the exponent $s$, the regularization of a singular
core, rotation, which can move the concentration point and nucleate vortices, and the
inverse-square endpoint.

\emph{What is known.} Minimizers of the attractive Gross--Pitaevskii (GP) functional
exist exactly below the Townes threshold~\cite{Weinstein1983,Seiringer}; as
$|g|\uparrow g_{*}$ they concentrate on the Townes profile at a rate set by the local
behavior of the trap~\cite{Seiringer,GuoZengZhou2016}, including singular
potentials~\cite{Phan2017}. In harmonic traps rotation below the centrifugal limit leaves
the threshold and the blow-up profile
unchanged~\cite{LNR2018,GuoLuoPeng2023,DinhNguyenRougerie,LeNguyen2026}. For a rotating
harmonic-plus-quartic trap whose rotating-frame potential vanishes on a ring, Guo, Li, and
Luo proved existence for every rotation speed, absence of vortices near threshold, and
concentration on the ring with an explicit rate and constant~\cite{GuoLiLuo2024}. An
attractive gas carries angular momentum by center-of-mass motion rather than by
vortices~\cite{WilkinGunnSmith1998,Lundh2004,Collin2005}: in rotating anharmonic traps
the lowest state breaks rotational symmetry and the cloud orbits the
axis~\cite{Lundh2004,Collin2005,KavoulakisJacksonBaym2004}, with a Goldstone mode in its
Bogoliubov spectrum~\cite{Collin2006}, and rotating attractive clouds collapse through
fragmentation~\cite{SaitoUeda2004}. For repulsive gases the TF and vortex asymptotics
are known in the slow- and fast-rotation
limits~\cite{IgnatMillot2006,CRY1,CRY2,BCPY2007,Nguyen2022,AftalionBlancNier2006}; in
harmonic-plus-quartic traps the cloud becomes annular and a giant vortex forms at fast
rotation~\cite{FetterJacksonStringari2005,KasamatsuTsubotaUeda2002,FuZaremba2006,Rougerie2013},
and weak-interaction rotational phase diagrams were recently mapped for power-law and
hard-wall traps~\cite{Kavoulakis2026}. In the inverse-square potential, cubic repulsion
arrests the quantum fall to the center in three dimensions but not in
two~\cite{SakaguchiMalomed2011,Malomed2018,Shamriz2020}, and many-body calculations in three
dimensions find a metastable gas above the collapsed state~\cite{AstrakharchikMalomed2015};
linear and nonlinear states in singular potentials and their regularization have recently
been reviewed and extended, mainly in three
dimensions~\cite{SakaguchiMalomed2026,SakaguchiMalomed2026b}.

\emph{This work.} We combine these results into one account of the dilation balance and
test it with converged radial solutions, a Bogoliubov calculation, and two-dimensional
imaginary-time states for both signs of $g$. Three results are new, to our knowledge.
(a)~\emph{When rotation moves the collapse.} The symmetry-broken state of
Refs.~\cite{Lundh2004,Collin2006,KavoulakisJacksonBaym2004} appears exactly when $\Omega$
exceeds the dipole frequency $\omega_1$ of the centered cloud, and near threshold
$\omega_1^2\propto\ell^{\,s-2}$: Kohn for $s=2$~\cite{Kohn1961,Dobson1994}, stiff for
$s<2$, soft for $s>2$ (\cref{sec:axis}). For $s=4$ the Townes trial gives
$\delta\simeq0.324\,\Omega^6$, which a Bogoliubov calculation confirms within $3\%$. On the
ring the width is set by the bare-trap curvature $\nabla^2V$, the power-law analog of the
Guo--Li--Luo law, and the same ring empties a repulsive TF cloud at
$\Omega_c=(3g/\pi)^{1/6}$; for $0<s<2$ there is no ring, and a rotating cloud of either
sign escapes unless boxed (\cref{sec:ring}).
(b)~\emph{The inverse-square endpoint.} The regularized energy is exactly covariant with
the cutoff, the condensate falls to the center for either sign of $g$, and repulsion only
adds a halo of density $(1+\kappa)/(2gr^2)$ and extent $\propto e^{g/[\pi(1+\kappa)]}$ for a
tail $-\kappa/(2r^2)$; the halo is universal, the core is not (\cref{sec:endpoint-theory}).
(c)~\emph{A numerical pitfall near collapse.} Discretized normalized gradient flows
converge to time-step-dependent states~\cite{Bao2004,ChenDongLiuXie2023,WangCaiSongWang2026};
for split-step imaginary time we derive the first-order modified operator and show that
near a collapse threshold its effect on the width grows as $1/\delta$ along the breathing
mode, invisible to a residual test (\cref{sec:dtau-num}).
Along the way we test known results quantitatively: the law $\ell\propto\delta^{1/(s+2)}$
for seven exponents, the cone law $\ell\propto\delta^{1/3}$ of a regularized core, which
is the $p=1$ case of Ref.~\cite{Seiringer}, and the TF scaling of repulsive condensates.
For repulsive vortex states we separate the TF energy error into its density-gradient and
flow parts. We do not claim as new the Townes threshold, the law $\delta^{1/(s+2)}$,
rotation independence in harmonic traps, the center-of-mass origin of the angular momentum
and the symmetry breaking it entails, the $\delta^{1/4}$ rate and its constant for the
harmonic-plus-quartic ring, the TF radius, the fall to the center itself, or the
time-step dependence of gradient-flow ground states. \Cref{tab:laws} lists the laws and
where they are tested. \Cref{sec:model,sec:theory} define the model and the scaling
structure, \cref{sec:conc,sec:rot,sec:endpoint} present nonrotating traps, rotation, and
the endpoint for both signs of $g$, and \cref{sec:exp} discusses experiments.

\section{Model and methods}
\label{sec:model}

\subsection{Gross--Pitaevskii model}
The order parameter $\psi(\x)$, $\x=(x,y)\in\R^2$, normalized as
$\|\psi\|_2^2=\int_{\R^2}|\psi(\x)|^2\,\dd\x=1$, minimizes the rotating-frame GP energy
\begin{equation}
  E[\psi]=\int_{\R^2}\Big[\half|\nabla\psi|^2+V|\psi|^2
  +\frac g2|\psi|^4-\Omega\,\psi^\ast\Lz\psi\Big]\,\dd\x ,
  \label{eq:E}
\end{equation}
where $r=|\x|$, $\Lz=-i(x\partial_y-y\partial_x)$, $\Omega\ge0$ is the
rotation frequency, and $g$ is real. The GP functional is the dilute limit of the
many-body problem~\cite{LSY1,LSY2D2001,LSe}, including the two-dimensional focusing
case~\cite{LNR2017note}. The trap is
\begin{equation}
  V_s(r)=\frac{r^s}{s},\qquad s\in[-2,0)\cup(0,4],
  \label{eq:V}
\end{equation}
so that $V_2=r^2/2$, $V_{-2}=-1/(2r^2)$ is the inverse-square endpoint, the radial force
$-r^{s-1}$ is attractive for every $s$, and $V_s<0$ for $s<0$. Units: for a trap
$\kappa r^s/s$ with $s\neq-2$ the length $\aho=(\hbar^2/m\kappa)^{1/(s+2)}$ and the energy
$\hbar^2/m\aho^2$ set $\hbar=m=\kappa=1$; for $s=2$ these are oscillator units. At $s=-2$
the strength $\kappa$ is dimensionless in units of $\hbar^2/m$; the computations use
$\kappa=1$, and \cref{sec:endpoint-theory} gives the dependence on $\kappa$. There is then
no intrinsic length, and lengths are quoted in a reference unit in which the box radius is
$\Rbox=2$. In a quasi-two-dimensional gas of $N$ atoms with axial length $a_z$ and
scattering length $a_s$, $g=\sqrt{8\pi}\,Na_s/a_z$ independently of
$s$~\cite{PetrovHolzmannShlyapnikov2000}. We measure the coupling from the Townes value
$g_{*}$ of \cref{eq:astar} by
\begin{equation}
  \delta=1+\frac{g}{g_{*}},
  \label{eq:delta}
\end{equation}
so that $\delta=1-|g|/g_{*}\in(0,1)$ for subcritical attraction, $\delta=1$ for the
ideal gas, and $\delta>1$ for repulsion.

For $s<0$ the singularity is regularized, $V_{s,\eps}(r)=(r+\eps)^s/s$ with cutoff
$\eps>0$, and outer confinement is provided by a disk of radius $\Rbox$. In the
analysis the disk carries a Dirichlet condition; in the computations it is the
stiff wall $V^{\rm wall}=\kappa_{\rm w}(r-\Rbox)_+^2$, $\kappa_{\rm w}=5000$, whose energy
$\int_{\R^2}V^{\rm wall}|\psi|^2\,\dd\x$ is added to \cref{eq:E} [$(\cdot)_+=\max(\cdot,0)$]. For $s>0$ a weak wall of the same form lies outside
the cloud. Completing the square in \cref{eq:E} with $\mathbf A=\Omega(-y,x)$ gives the magnetic form
\begin{equation}
  E[\psi]=\int_{\R^2}\Big[\half|(-i\nabla-\mathbf A)\psi|^2+\Veff|\psi|^2
  +\frac g2|\psi|^4\Big]\,\dd\x
  \label{eq:Emag}
\end{equation}
with the rotating-frame potential $\Veff=V-\half\Omega^2r^2$.
Stationary states solve $H_\Omega\psi=\mu\psi$ with the GP operator
$H_\Omega=-\half\nabla^2+V+V^{\rm wall}+g|\psi|^2-\Omega\Lz$ and the chemical potential
$\mu=\avg{\psi,H_\Omega\psi}$, where $\avg{f_1,f_2}=\int_{\R^2}f_1^\ast(\x)f_2(\x)\,\dd\x$;
expectation values in the state $\psi$ are written $\avg{A}=\avg{\psi,A\psi}$.

\subsection{Numerical methods}
\label{sec:methods}
Four independent tools are used; details and convergence data are in
\cref{app:numerics}.

\emph{Two-dimensional states.} All rotating and nonrotating two-dimensional states
are computed by normalized imaginary-time propagation~\cite{Bao2004,Antoine2014} in
the imaginary time $\tau$ with step $\Delta\tau$ and the Strang
splitting~\cite{Strang1968}
\begin{equation}
  \psi\mapsto\mathcal N\big[e^{-\Delta\tau U'/2}P_2(\Omega\Delta\tau\Lz)
  \mathcal F^{-1}e^{-\Delta\tau k^2/2}\mathcal F e^{-\Delta\tau U/2}\psi\big].
  \label{eq:split}
\end{equation}
Here $\mathcal F$ is the discrete Fourier transform with wave vector $\mathbf k$,
$P_2(a\Lz)=1+a\Lz+a^2\Lz^2/2$ applies the rotation $e^{a\Lz}$ with spectral derivatives
and a local error $O(\Delta\tau^3)$ ($\Lz$ commutes with the Laplacian, so the kinetic and
rotation factors need no splitting), $\mathcal N$ restores the norm, $U=V+V^{\rm wall}+g|\psi|^2$ is evaluated on the state entering the step,
and $U'$ on the intermediate state entering the second half step. The grids
($N_{\rm g}^2$ points on a square of side $\Lsq$, spacing $\Delta x=\Lsq/N_{\rm g}$),
boxes, and regulators are listed in \cref{tab:grids}. A state is accepted when its
energy is stationary and the relative residual
\begin{equation}
  \etaGP=\frac{\|(H_\Omega-\mu)\psi\|_2}{\max\{\|H_\Omega\psi\|_2,|\mu|\}}
  \label{eq:res}
\end{equation}
meets the predesignated gate of \cref{tab:grids}; states that are energy
stationary but miss the gate are retained, flagged, and not used for quantitative
conclusions unless stated. Rotating attractive states in the quartic trap are
started from a centered and a displaced initial state, and the lower energy is kept;
we refer to it as the lowest state found. Repulsive rotating states start from the TF
density carrying a triangular vortex lattice of areal density $\Omega/\pi$, the
coarse-grained density of rigid rotation, restricted to the TF support; they form the
stationary branch selected by this initial state and are not certified ground states
(\cref{sec:rot-rep}). One unmodified version of the
solver is used for every state (\cref{app:numerics}).

\begin{table}[t]
\caption{Two-dimensional computations: sign of $g$, exponents, grid points per side
$N_{\rm g}$, side $\Lsq$ of the computational square, box radius $\Rbox$, regulator $\eps$,
and residual gate on $\etaGP$; the gate is $0.05$ whenever $s\le-1.9$. For $g>0$, $s>0$ the
square is adapted to the size of the cloud.}
\label{tab:grids}
\setlength{\tabcolsep}{3pt}
\begin{ruledtabular}
\begin{tabular}{cccccc}
$g$ & $s$ & $N_{\rm g}$ & $\Lsq$ & $\Rbox$; $\eps$ & gate\\ \hline
$>0$ & $>0$ & 256 & adapted & --- & 0.02\\
$>0$ & $<0$ & 512 & 20 & 7; 0.02 & 0.02\\
$<0$ & $>0$ & 512 & 12 & --- & 0.002\\
$<0$ & $<0$ & 1024 & 12 & 5; 0.03, 0.06 & 0.002\\
both & $[-2,-1.5]$ & 256--1024 & 5 & 2; 0.025--0.25 & 0.02\\
\end{tabular}
\end{ruledtabular}
\end{table}

\emph{Radial references.} For $\Omega=0$ the trap and wall increase with $r$, so the
ground state is radial (symmetric decreasing rearrangement). Every nonrotating
problem is therefore solved a second time as a radial problem with the identical
potentials: cell-centered finite volumes on a graded grid, cell-averaged
potentials, a backward-Euler gradient flow followed by a bordered Newton
iteration, and Richardson extrapolation over three nested grids (observed order
$2.00$, residual $\le10^{-8}$). The radial values carry relative uncertainties below
$2\times10^{-4}$ and reach regulators and couplings outside the two-dimensional range.

\emph{Bogoliubov spectrum of centered states.} The rotating-frame energy Hessian of
a centered radial state $\psi_0(r)>0$ separates into angular-momentum sectors $m$.
In each sector it reduces to the radial operators
\begin{equation}
  L_\pm^{(m)}=-\half\Big(\partial_r^2+\frac1r\partial_r-\frac{m^2}{r^2}\Big)
  +V-\mu+c_\pm\,g\,\psi_0^2,
  \label{eq:Lpm}
\end{equation}
with $c_+=3$ and $c_-=1$,
whose product $L_-^{(m)}L_+^{(m)}$ has the squared Bogoliubov frequencies $\omega_m^2$
of the nonrotating state as eigenvalues (\cref{sec:axis}). We compute the lowest
$\omega_m$ on uniform radial grids with second-order finite differences,
Richardson-extrapolated over three grids (\cref{app:bogo}). In a harmonic trap the
dipole frequency reproduces the Kohn value $\omega_1=1$ to $3\times10^{-7}$ for
$g=-3$, $0$, and $20$.

\emph{Variational bound.} For every attractive point we also minimize the energy
exactly over the gauge-boosted Townes trial of \cref{sec:theory}, with the actual
trap and $\Omega$. This gives a variational energy $\Evar$, which bounds the ground-state
energy from above, and its minimizing width $\lvar$ and center $x_{0,\rm var}$. A
computed state with $E>\Evar$ is thereby excluded as a ground state.

\section{Scaling structure}
\label{sec:theory}

\Cref{sec:conc-theory} restates known results for both signs of $g$ and writes the TF energy
error exactly; \cref{sec:cone} applies a known rate to regularized cores;
\cref{sec:axis,sec:ring,sec:endpoint-theory} are new. Each statement is identified as exact,
as an identity within the Townes trial family, or as an asymptotic estimate.

\subsection{One dilation, two signs}
\label{sec:conc-theory}
Let $Q$ be the positive radial solution of $-\nabla^2Q+Q-Q^3=0$ on $\R^2$ with
$Q(r)\to0$ as $r\to\infty$. With $\|f\|_p=(\int_{\R^2}|f(\x)|^p\,\dd\x)^{1/p}$, the sharp
Gagliardo--Nirenberg inequality~\cite{Weinstein1983}
\begin{equation}
  \|\psi\|_4^4\le\frac{2}{\|Q\|_2^2}\,\|\nabla\psi\|_2^2\,\|\psi\|_2^2
  \label{eq:GNineq}
\end{equation}
gives for $g<0$ and $\|\psi\|_2=1$
\begin{equation}
  \half\|\nabla\psi\|_2^2+\frac g2\|\psi\|_4^4\ge\frac{\delta}{2}\,\|\nabla\psi\|_2^2,
  \label{eq:GN}
\end{equation}
where $\delta=1-|g|/g_{*}$ [\cref{eq:delta}] and
\begin{equation}
  g_{*}=\half\|Q\|_2^2=5.850448,
  \label{eq:astar}
\end{equation}
with $\|Q\|_2^2=11.700897$ from shooting. For $-2<s\le4$ the trap is
either bounded below or infinitesimally form bounded with respect to $-\nabla^2$, and the
diamagnetic inequality extends the bound to $\Omega>0$. Hence, for $g<0$, minimizers exist
for every $|g|<g_{*}$, every exponent and regulator, and every $\Omega$ for which $\Veff$
or the box confines~\cite{Seiringer,Phan2017,LNR2018,GuoLuoPeng2023,GuoLiLuo2024}. For
$|g|>g_{*}$ the energy is unbounded below. At $|g|=g_{*}$ it is unbounded below for an
unregularized singular trap, whereas for $\Omega=0$ and a trap bounded below the infimum
equals $\min V$ and is not attained~\cite{Seiringer}. Trap shape and rotation do not move
the threshold. In a Gaussian trial space rotation delays the collapse to stronger
attraction~\cite{KavoulakisJacksonBaym2004}; in the full functional the angular momentum can
be carried by the center of mass of a concentrating Townes profile at finite cost, so the
shift disappears.

They do control the approach to it. With the normalized Townes trial of width $\ell$ and
the Townes average,
\begin{equation}
  Q_\ell(\x)=\frac{Q(\x/\ell)}{\ell\|Q\|_2},\qquad
  \avg{f}_Q=\frac{\int_{\R^2}f(\y)Q(\y)^2\,\dd\y}{\int_{\R^2}Q(\y)^2\,\dd\y},
  \label{eq:trial}
\end{equation}
the Pohozaev identities $\|\nabla Q\|_2^2=\|Q\|_2^2=\half\|Q\|_4^4$ give the trial energy
\begin{equation}
  E(\ell)=\frac{\delta}{2\ell^2}+\frac{m_s}{s}\,\ell^s,\qquad m_s=\avg{|\y|^s}_Q .
  \label{eq:Eell}
\end{equation}
For every $s\in(-2,0)\cup(0,\infty)$ its global minimum is at
\begin{equation}
  \ell_\ast=\Big(\frac{\delta}{m_s}\Big)^{1/(s+2)},\qquad
  e_\ast=\frac{s+2}{2s}\,m_s\,\ell_\ast^{\,s}
  \label{eq:conc}
\end{equation}
(\cref{app:lean}). For $s>0$ this is the leading-order concentration law of
Refs.~\cite{Seiringer,GuoZengZhou2016}; for $-2<s<0$ it agrees with the rigorous
asymptotics for singular potentials~\cite{Phan2017}. As $s\to-2^+$ both the exponent
$1/(s+2)$ and $m_s$ diverge, with $(s+2)\,m_s\to2\pi Q(0)^2/\|Q\|_2^2=2.614$
(\cref{tab:townes}): the law loses its scale before the threshold moves.

\begin{table}[t]
\caption{Moments $m_s=\avg{|\y|^s}_Q$ of the Townes profile [\cref{eq:Eell}], from
shooting.}
\label{tab:townes}
\begin{ruledtabular}
\begin{tabular}{ccccc}
$s$ & 4 & 3 & 2 & 1\\ \hline
$m_s$ & 4.3398 & 2.0279 & 1.1875 & 0.9132\\ \hline
$s$ & $-0.5$ & $-1$ & $-1.5$ & $-1.9$\\ \hline
$m_s$ & 1.2587 & 1.9217 & 4.2607 & 24.860\\
\end{tabular}
\end{ruledtabular}
\end{table}

For $g>0$ the same balance runs the other way. Once the cloud is large, the contact energy
dominates the kinetic one and the TF density $n^{\rm TF}=(\mu^{\rm TF}-V)_+/g$ holds. For
$s>0$ and $\Omega=0$ its radius and chemical potential are
\begin{equation}
  R^{\rm TF}_0=\Big[\frac{g(s+2)}{\pi}\Big]^{\frac1{s+2}},\qquad
  \mu^{\rm TF}_0=\frac1s\Big[\frac{g(s+2)}{\pi}\Big]^{\frac s{s+2}}
  \label{eq:RTF}
\end{equation}
(\cref{app:tf}); the subscript $0$ denotes $\Omega=0$. The exponent is that of
\cref{eq:conc} with $g$ in place of $\delta$: in both cases a cloud of size $L$ balances a
$\lambda^2$ coefficient against the trap, $L^{s+2}\propto\delta$ for attraction and
$L^{s+2}\propto g$ for repulsion. The difference is where the trap acts. A concentrating
cloud samples the trap only near one point; a TF cloud samples it over its whole support,
and the kinetic energy is a boundary-layer correction controlled by the healing length
$\xi=(2gn)^{-1/2}$, with $\xi/R^{\rm TF}_0\propto g^{-1/2}$ for every $s>0$. Rotation
competes with the trap when $\Omega^2(R^{\rm TF}_0)^2\sim\mu^{\rm TF}_0$, i.e., at fixed
reduced rotation $\zeta=\Omega\,g^{-(s-2)/[2(s+2)]}$, the two-dimensional analog of the
reduced rotation of Ref.~\cite{BCPY2007}.

Under rotation the phase of a repulsive state carries vortices, and the kinetic energy is
no longer small. Writing $\psi=|\psi|e^{i\varphi}$ splits it as
$|\nabla\psi|^2=|\nabla|\psi||^2+|\psi|^2|\nabla\varphi|^2$~\cite{Madelung1927}, and
completing the square with the rotation term gives the exact identity
\begin{align}
  E[\psi]=\int_{\R^2}\Big[&\half|\nabla|\psi||^2
  +\half|\psi|^2\big|\nabla\varphi-\Omega\hat{\mathbf z}\times\x\big|^2\nonumber\\
  &+\Veff|\psi|^2+\frac g2|\psi|^4\Big]\,\dd\x ,
  \label{eq:Esplit}
\end{align}
with $\hat{\mathbf z}$ the unit normal to the plane ($V$ includes the wall when present).
The rotating-frame TF functional $E_{\rm TF}[n]=\int_{\R^2}(\Veff n+\frac g2n^2)\,\dd\x$,
whose minimizer is $(\mu^{\rm TF}-\Veff)_+/g$~\cite{Nguyen2022}, keeps the last two terms.
It omits two nonnegative energies: the density-gradient energy, and the mismatch between
the flow and rigid rotation, which for a vortex lattice is the vortex energy beyond the
coarse-grained rigid flow~\cite{Fetter2009}. We measure them separately,
\begin{align}
  \epsgrad&=\frac{1}{E_{\rm int}}\int_{\R^2}\half|\nabla|\psi||^2\,\dd\x,\nonumber\\
  \gflow&=\frac{1}{E_{\rm int}}\int_{\R^2}\half|\psi|^2\big|\nabla\varphi-\Omega\hat{\mathbf z}\times\x\big|^2\,\dd\x,
  \label{eq:epsgrad}
\end{align}
with $E_{\rm int}=\frac g2\int_{\R^2}|\psi|^4\,\dd\x$, so that
$E[\psi]-E_{\rm TF}[|\psi|^2]=(\epsgrad+\gflow)E_{\rm int}$. Neither ratio counts the rigid
part of the circulation, which dominates the total kinetic energy of a vortex
lattice~\cite{Nore1997,Bradley2012}; $\gflow$ follows from recorded quantities through
$\gflow E_{\rm int}=E_{\rm kin}-\epsgrad E_{\rm int}-\Omega\avg{\Lz}+\half\Omega^2\avg{r^2}$.

\subsection{Regularized singular cores: the cone law}
\label{sec:cone}
That the attractive law depends only on the local trap shape has an immediate consequence
for regularized singular traps. At
fixed $\eps$, $V_{s,\eps}(r)=\eps^s/s+\eps^{s-1}r+O(r^2)$: the regularized core is a cone.
Once $\ell\ll\eps$ the trial energy is therefore
$E(\ell)\simeq V_{s,\eps}(0)+\delta/(2\ell^2)+m_1\eps^{s-1}\ell$, and
\begin{equation}
  \ell_\ast^3\simeq\frac{\delta}{m_1\,\eps^{\,s-1}},\qquad
  \delta\ll m_1\eps^{\,s+2},
  \label{eq:cone}
\end{equation}
the $s=1$ law with a cutoff-dependent prefactor. This is the $p=1$ case of the rate
$(g_{*}-|g|)^{1/(p+2)}$ proved in Ref.~\cite{Seiringer} for traps that vanish like $|\x|^p$
at their minimum; what the regulator changes is which exponent applies. The unregularized
law $\delta^{1/(s+2)}$ holds only while $\ell_\ast\gg\eps$, and in between lies a crossover.
A smooth core with a quadratic minimum would give $\ell\propto\delta^{1/4}$ instead: the
exponent near threshold belongs to the shape of the core, not to the far tail.

\subsection{Leaving the axis: the dipole criterion}
\label{sec:axis}
\emph{Exact criterion.} Let $\psi_0(r)>0$ be the centered stationary state, which does not
depend on $\Omega$ because $\Lz\psi_0=0$. Writing a perturbation of $\psi_0$ as $a+ib$ with
real $a,b$, the second variation of the rotating-frame energy at fixed norm is
\begin{equation}
  \delta^2E=\avg{a,L_+a}+\avg{b,L_-b}-2\Omega\int_{\R^2}a\,\partial_\theta b\,\dd\x ,
  \label{eq:hessian}
\end{equation}
with $L_\pm=-\half\nabla^2+V-\mu+c_\pm g\psi_0^2$ and $\theta$ the polar angle. In the
sector $a=\alpha(r)\cos m\theta$, $b=\beta(r)\sin m\theta$, $m\ge1$, this becomes a
$2\times2$ radial form with diagonal $L_\pm^{(m)}$ of \cref{eq:Lpm} and off-diagonal
$-m\Omega$. It is singular exactly when $(m\Omega)^2$ is an eigenvalue of
$L_-^{(m)}L_+^{(m)}$, i.e., a squared Bogoliubov frequency $\omega_m^2$ of the nonrotating
state. The centered state is therefore a local minimum of the rotating-frame energy if and
only if $\Omega<\min_{m\ge1}\omega_m/m$ (the operators $L_\pm^{(m)}$ are positive for
$m\ge1$ in all cases computed), the Landau criterion for a rotating
trap~\cite{Pitaevskii2003}. For the near-threshold states considered here the minimum is
attained at $m=1$ (\cref{sec:rot}), so the centered cloud becomes unstable when $\Omega$
exceeds its dipole frequency $\omega_1(\delta)$. The criterion is energetic: beyond it the
centered state remains stationary, and an isotropic cloud leaves the axis only through
dissipation or a perturbation that breaks rotational symmetry. It locates the symmetry
breaking seen in the anharmonic-trap calculations of
Refs.~\cite{Lundh2004,Collin2006,KavoulakisJacksonBaym2004}, whose zero-frequency
mode~\cite{Collin2006} is the rotation of the displaced cloud about the axis.

\emph{Townes trial.} A rigid, boosted displacement of $Q_\ell$ by $\x_0$ (see
\cref{sec:ring}) changes the trial energy by
$\half x_0^2[\avg{\partial_x^2V(\ell\y)}_Q-\Omega^2]+O(x_0^4)$, with $x_0=|\x_0|$. The
trial dipole frequency is therefore
\begin{equation}
  \omega_{\rm var}^2=\avg{\partial_x^2V(\ell\y)}_Q
  =\frac s2\,m_{s-2}\,\ell^{\,s-2}\qquad(s>0),
  \label{eq:wvar}
\end{equation}
where isotropy gives $\avg{\partial_x^2V}_Q=\half\avg{\nabla^2V}_Q$ and
$\nabla^2V_s=s\,r^{s-2}$. It equals the Kohn value $1$ for $s=2$, diverges as $\ell\to0$ for
$0<s<2$, and vanishes as $\ell\to0$ for $s>2$. Only for $s>2$ can rotation move a
collapsing cloud off the axis, and there every $\Omega>0$ eventually does. With
$\ell=\ell_\ast$ of \cref{eq:conc}, the trial threshold is
\begin{equation}
  \dvar(\Omega)=m_s\Big(\frac{2\Omega^2}{s\,m_{s-2}}\Big)^{\frac{s+2}{s-2}},
  \label{eq:dcvar}
\end{equation}
i.e., $\dvar=m_4(\Omega^2/2m_2)^3=0.3239\,\Omega^6$ for $s=4$ and $\dvar=0.4204\,\Omega^{10}$
for $s=3$.

\emph{Quartic trap.} For $s=4$ the trial energy is available in closed form along the whole
displaced branch. From the isotropic moments of $Q^2$,
\begin{equation}
  \avg{|\y|^2}_Q=m_2,\quad \avg{(\mathbf e\cdot\y)^2}_Q=\frac{m_2}{2},\quad
  \avg{|\y|^4}_Q=m_4,
  \label{eq:moments}
\end{equation}
for any unit vector $\mathbf e$, one finds
$\avg{|\x_0+\ell\y|^4}_Q=x_0^4+4m_2\ell^2x_0^2+m_4\ell^4$, and the trial energy of the
boosted lump (\cref{sec:ring}) is
\begin{equation}
  E(\ell,x_0)=\frac{\delta}{2\ell^2}+\frac{m_4}{4}\ell^4+m_2\ell^2x_0^2
  +\frac{x_0^4}{4}-\frac{\Omega^2x_0^2}{2},
  \label{eq:quartic}
\end{equation}
minimized over the center by $x_0^2=(\Omega^2-2m_2\ell^2)_+$. On the displaced branch the
reduced energy has the negative quartic coefficient $(m_4-4m_2^2)/4$, but its second
derivative at the branch point is $(6m_4-8m_2^2)\ell^2>0$, and its critical points are
unique (\cref{app:lean}). The transition at $\dvar$ is therefore continuous; because of the
rotational symmetry the displaced states form a circle, and the transition is a pitchfork
in any fixed radial direction. Below $\dvar$ the center moves out as
$x_0^2=\Omega^2-2m_2\ell^2$ and reaches the ring $r_0=\Omega$ of $\Veff$ only as $\ell\to0$.

\subsection{The ring of the rotating-frame potential}
\label{sec:ring}
For $s>2$ and $\Omega>0$, $\Veff=V-\half\Omega^2r^2$ has a minimum ring at
$r_0=\Omega^{2/(s-2)}$, with $\Veff(r_0)=-\tfrac{s-2}{2s}\,\Omega^{2s/(s-2)}$. Both signs of
$g$ end up on it at fast rotation.

\emph{Attraction: a boosted lump.} For $\psi=e^{i\mathbf p\cdot\x}Q_\ell(\x-\x_0)$ with
internal coordinate $\y$ defined by $\x=\x_0+\ell\y$, the boost that minimizes the energy
is $\mathbf p=\Omega\hat{\mathbf z}\times\x_0$. The rotating-frame energy of this trial is
exactly
\begin{equation}
  E(\ell,\x_0)=\frac{\delta}{2\ell^2}+\avg{V(|\x_0+\ell\y|)}_Q-\half\Omega^2x_0^2 ,
  \label{eq:Eoff}
\end{equation}
and $\avg{\Lz}=\Omega x_0^2$, the angular momentum of rigid center-of-mass rotation
(\cref{app:trial}). The centrifugal potential acts on the center but not on the internal
coordinate: in the magnetic form \cref{eq:Emag}, the term $-\half\Omega^2\ell^2|\y|^2$ of
$\Veff$ is cancelled by the magnetic kinetic energy of the boosted lump, the Galilean
covariance of the rotating-frame functional. Expanding \cref{eq:Eoff} about $r_0$ gives
\begin{align}
  \ell_\ast^4&=\frac{2\delta}{m_2\nabla^2V(r_0)}=\frac{2\delta}{s\,\Omega^2m_2},\nonumber\\
  e_\ast&=\Veff(r_0)+\Omega\Big(\frac{s\,m_2\,\delta}{2}\Big)^{1/2}.
  \label{eq:ring}
\end{align}
The curvature that enters is $\nabla^2V(r_0)=s\Omega^2$, not
$\nabla^2\Veff(r_0)=(s-2)\Omega^2$; the latter would overestimate $\ell_\ast$ by
$[s/(s-2)]^{1/4}$, a factor $1.19$ for $s=4$. The first form of \cref{eq:ring} holds for any
trap with a ring minimum of $\Veff$. For the harmonic-plus-quartic trap of
Ref.~\cite{GuoLiLuo2024} it reproduces the rigorous rate and constant of that work
(\cref{app:trial}); the power-law family, in which the ring radius moves with $\Omega$, is
the case treated here.

\emph{Repulsion: an annulus.} The rotating-frame TF density $(\mu^{\rm TF}-\Veff)_+/g$ loses
its center when $\mu^{\rm TF}$ falls below $\Veff(0)=0$, and beyond that point it is an
annulus about $r_0$. For $s=4$ the normalization at $\mu^{\rm TF}=0$ gives
\begin{equation}
  \Omega_c=\Big(\frac{3g}{\pi}\Big)^{1/6},\qquad
  \zeta_c=\Omega_c\,g^{-1/6}=\Big(\frac3\pi\Big)^{1/6}=0.992
  \label{eq:Omegac}
\end{equation}
(\cref{app:tf}), the pure-quartic member of the harmonic-plus-quartic analyses of
Refs.~\cite{FetterJacksonStringari2005,KasamatsuTsubotaUeda2002}; $\zeta_c$ is the reduced
rotation $\zeta$ at $s=4$. The threshold marks a change of the TF support, not the onset of
a giant vortex, which in GP requires the annulus to expel its inner vortices at larger
$\Omega$~\cite{FuZaremba2006,Rougerie2013}. The two signs share the ring but not its
physics: the lump sees the curvature of the bare trap at $r_0$, the annulus sees $\Veff$
over its whole width.

\emph{No ring for $s<2$.} For $0<s<2$ and $\Omega>0$, $\Veff\to-\infty$ at large $r$. The
lowest-Landau-level~\cite{Ho2001,Cooper2008} ring $\psi_m\propto z^me^{-\Omega|z|^2/2}$, with
$z=x+iy$ and rms radius
$R_m=[(m+1)/\Omega]^{1/2}$, has magnetic kinetic energy $\Omega$; by Jensen's inequality
$\avg{r^s}\le R_m^s$, so
\begin{equation}
  E[\psi_m]\le\Omega+\Veff(R_m)+O(m^{-1/2})\to-\infty\qquad(m\to\infty)
  \label{eq:LLL}
\end{equation}
for every $g\ge0$, and a fortiori for $g<0$. The rotating problem for $0<s<2$, and likewise
for $s<0$, has no ground state without an outer box: the cloud escapes to large radius,
which is why all rotating singular-trap states are boxed. In a boxed cloud the net winding
of the phase on a closed contour on which $\psi\neq0$ is the degree of $\psi/|\psi|$, an
integer that stays constant along any continuous branch of states as long as no zero of
$\psi$ crosses the contour.

\subsection{The inverse-square endpoint}
\label{sec:endpoint-theory}
At $s=-2$ the regularized trap $V_\eps(r)=-\half(r+\eps)^{-2}$ is $\eps^{-2}W(r/\eps)$ with
$W(u)=-\half(1+u)^{-2}$. Under a simultaneous dilation of $r$ and $\eps$ it scales like the
kinetic and contact energies, and the regularized problem is exactly covariant.

\emph{Proposition 1.} Let $E_{\eps,R}(g)$ be the GP ground-state energy for $V_\eps$ on the
disk of radius $R$ ($R=\Rbox$ in the computations) with a Dirichlet condition, $\Omega=0$,
and $g>-g_{*}$. For $\Lambda>0$ let $D_\Lambda$ be the disk of radius $\Lambda$ in the
rescaled variable $\uu=\x/\eps\in\R^2$, $u=|\uu|$, and let $e_\Lambda(g)$ be the infimum of
\begin{equation}
  \cE_g[\phi]=\int_{D_\Lambda}\Big[\half|\nabla\phi|^2+W(u)|\phi|^2
  +\frac g2|\phi|^4\Big]\,\dd\uu
  \label{eq:resc}
\end{equation}
over $\phi$ in the Sobolev space $H_0^1(D_\Lambda)$ with $\|\phi\|_2=1$. Then (i)
$E_{\eps,R}(g)=\eps^{-2}e_{R/\eps}(g)$; (ii) $e_\Lambda(g)$ is nonincreasing in $\Lambda$ and
converges to the infimum $e(g)$ over $\R^2$, with $e(g)\ge-\half$, and $e(g)\ge-\pi/(24g)$
for $g>0$; (iii) $e(g)<0$ for every $g>-g_{*}$. Hence $\eps^2E_{\eps,R}(g)\to e(g)<0$ as
$\eps\to0$: the energy falls as $-\eps^{-2}$ for attractive, ideal, and repulsive gases
alike.

\emph{Proof.} (i) The unitary dilation $\psi(\x)=\eps^{-1}\phi(\x/\eps)$ maps the disk of
radius $R$ onto $D_{R/\eps}$, and each term acquires the factor $\eps^{-2}$. (ii) Extension
by zero shows that $e_\Lambda$ is nonincreasing. Since $W\ge-\half$, the contact term is
nonnegative for $g\ge0$ and controlled by \cref{eq:GN} for $-g_{*}<g<0$, so
$\cE_g\ge-\half$. For $g>0$, Young's inequality $\tfrac g2|\phi|^4+W|\phi|^2\ge-W^2/(2g)$
and $\int_{\R^2}W(|\uu|)^2\,\dd\uu=\pi/12$ give $\cE_g\ge-\pi/(24g)$. (iii) Take
$\phi_\rho(\uu)=\rho^{-1}\chi_0(\uu/\rho)$ with a smooth, normalized, radial
$\chi_0\in C_c^\infty(\R^2)$, $\chi_0(0)>0$. Its kinetic and contact energies are
$(A+gB)/\rho^2$, with $A=\half\|\nabla\chi_0\|_2^2$ and $B=\half\|\chi_0\|_4^4$, while its
trap energy is
\begin{equation}
  -\frac1{\rho^2}\int_{\R^2}\frac{|\chi_0(\vv)|^2}{2(|\vv|+\rho^{-1})^2}\,\dd\vv
  \le-\frac{\nu_1}{\rho^2}\,(\ln\rho-\nu_2),
  \label{eq:logtrial}
\end{equation}
with constants $\nu_1,\nu_2>0$, because the two-dimensional integral of $|\vv|^{-2}$ diverges
logarithmically at the origin. Thus $\cE_g[\phi_\rho]<0$ for large $\rho$. $\square$

Statement (iii) is the fall to the center in the $1/r^2$
potential~\cite{Case1950,LandauLifshitzQM}, which in two dimensions occurs for every
coupling because the Hardy constant vanishes; that cubic repulsion does not arrest it in
two dimensions is known~\cite{Malomed2018,Shamriz2020}. What Proposition~1 adds is the
covariance (i): the entire cutoff dependence of the regularized Dirichlet problem is
carried by one function $e_\Lambda(g)$ of the single ratio $\Lambda=\Rbox/\eps$. Three
remarks fix its scope. First, the proposition concerns energies; that the minimizing
profiles converge as $\Lambda\to\infty$ is a numerical finding for the couplings studied
(\cref{sec:endpoint}), and we denote by $\phi_g$ the radial profile so obtained. A proof
would follow by concentration compactness~\cite{Lions1984} from the strict binding
inequality, which we have not established. Second, for the soft wall used in the
computations $\eps^2E$ depends in addition on $\kappa_{\rm w}\eps^4$; the covariance is exact
only for the Dirichlet disk and holds approximately when the profile stays away from the
wall (\cref{app:endpoint}). Third, any regularization $V_\eps=\eps^{-2}W(r/\eps)$ with
$W(u)\simeq-\half u^{-2}$ at large $u$ has the same covariance with a different function
$e_\Lambda(g)$; the $W$ used here is the one above unless stated otherwise.

\emph{Core and halo (asymptotic, $g\gg1$).} The trial of (iii) already shows how repulsion
enters: for a nearly flat profile of radius $\rho$ one has $B\simeq1/(2\pi)$ and
$\nu_1\simeq1$, so the energy is $(K_0+g/2\pi-\ln\rho)/\rho^2$ with $K_0=O(1)$ independent
of $g$; it is minimal at $\ln\rho=K_0+g/2\pi+\half$ (\cref{app:lean}), and the cloud grows as
$\ln\rho\simeq g/2\pi$. The stationary equation sharpens this. For $g>0$ write
$\phi=\chi/\sqrt g$, so that
\begin{equation}
  g\,\cE_g=\int_{D_\Lambda}\Big[\half|\nabla\chi|^2+W(u)\chi^2+\half\chi^4\Big]\,\dd\uu,
  \qquad \|\chi\|_2^2=g .
  \label{eq:chi}
\end{equation}
Far from the core, where $W\simeq-\half u^{-2}$ and the chemical potential is negligible,
the stationary equation $-\half\nabla^2\chi-\chi/(2u^2)+\chi^3=0$ has the solution
$\chi=1/u$, since $\nabla^2u^{-1}=u^{-3}$ in two dimensions. The profile is therefore a core
of mass $O(1/g)$ dressed by a halo
\begin{equation}
  |\phi_g(u)|^2\simeq\frac{1}{g\,u^2},\qquad 1\ll u\lesssim L_h(g) .
  \label{eq:halo}
\end{equation}
In three dimensions the same balance produces a density $\propto r^{-2}$ near the origin,
which is integrable there and allows repulsion to arrest the
fall~\cite{SakaguchiMalomed2011}. In two dimensions
$\int_{a<|\uu|<b}|\uu|^{-2}\,\dd\uu=2\pi\ln(b/a)$ diverges logarithmically at both ends: at
small $u$ this is why repulsion cannot stop the fall, and with a cutoff it fixes the size of
the halo through the normalization $(2\pi/g)\ln L_h\simeq1$, i.e.,
\begin{equation}
  L_h(g)\simeq C\,e^{g/2\pi} .
  \label{eq:Lh}
\end{equation}
Here $C$ absorbs the core mass and the cutoff of the halo by the small chemical potential.
Because the $1/u^2$ halo dominates the second moment, we define the halo extent
operationally through the rms radius,
\begin{align}
  L(g)&\equiv\frac{r_{\rm rms}}{\eps}
  =\Big[\int_{\R^2}|\uu|^2|\phi_g(\uu)|^2\,\dd\uu\Big]^{1/2},\nonumber\\
  L_h(g)&\equiv\Big(\frac g\pi\Big)^{1/2}L(g),
  \label{eq:Ldef}
\end{align}
which for a sharply cut halo of radius $L_h$ is exact. The halo \cref{eq:halo} and the
coefficient $1/2\pi$ in \cref{eq:Lh} depend only on the tail of $W$; the constant $C$, the
core mass, and $e(g)$ depend on the regularization.

\emph{Tail strength.} For a tail $-\kappa/(2r^2)$, i.e., $V_\eps=\eps^{-2}\kappa W(r/\eps)$,
Proposition~1 holds with the bounds $e(g)\ge-\kappa/2$ and $e(g)\ge-\pi\kappa^2/(24g)$, and the
same argument gives $\chi=[(1+\kappa)/2]^{1/2}/u$, i.e.,
\begin{equation}
  |\phi_g(u)|^2\simeq\frac{1+\kappa}{2g\,u^2},\qquad
  \ln L_h(g)\simeq\frac{g}{\pi(1+\kappa)}+\ln C ,
  \label{eq:halokappa}
\end{equation}
with $L_h\equiv[2g/\pi(1+\kappa)]^{1/2}L(g)$; \cref{eq:halo,eq:Lh,eq:Ldef} are the case
$\kappa=1$, which the computations use unless stated otherwise. In the original coordinate
$r$ the halo is $|\psi|^2\simeq(1+\kappa)/(2gr^2)$ for $\eps\ll r\lesssim\eps L_h(g)$, i.e., an
atom density $N|\psi|^2\simeq(1+\kappa)\,a_z/(2\sqrt{8\pi}\,a_s r^2)$ that does not depend
on $N$.

\begin{table*}[t]
\caption{The scaling laws of the power-law family and where they are tested; ``known''
marks results that we test rather than claim.}
\label{tab:laws}
\begin{ruledtabular}
\begin{tabular}{llll}
Regime & Law & Status & Tested in\\ \hline
\tcell{3.3cm}{Attraction, $\Omega=0$} & \tcell{6.5cm}{threshold $g_{*}$; $\ell_\ast=(\delta/m_s)^{1/(s+2)}$} & \tcell{4.3cm}{known~\cite{Seiringer,GuoZengZhou2016,Phan2017}} & \cref{sec:conc-att}\\
\tcell{3.3cm}{Regularized core, $\ell\ll\eps$} & \tcell{6.5cm}{$\ell_\ast^3\simeq\delta/(m_1\eps^{\,s-1})$} & \tcell{4.3cm}{$p=1$ case of Ref.~\cite{Seiringer}} & \cref{sec:cone-num}\\
\tcell{3.3cm}{Repulsion, $\Omega=0$, $s>0$} & \tcell{6.5cm}{$R^{\rm TF}_0=[g(s+2)/\pi]^{1/(s+2)}$} & \tcell{4.3cm}{known} & \cref{sec:tf-num}\\
\tcell{3.3cm}{Attraction, rotating, $s>2$} & \tcell{6.5cm}{leaves the axis for $\Omega>\omega_1$,   $\omega_1^2\propto\ell^{\,s-2}$; $\dGP\simeq0.32\,\Omega^6$ for $s=4$} & \tcell{4.3cm}{new} & \cref{sec:rot-axis}\\
\tcell{3.3cm}{Attraction on the ring} & \tcell{6.5cm}{$\ell_\ast^4=2\delta/[m_2\nabla^2V(r_0)]$} & \tcell{4.3cm}{power-law analog of Ref.~\cite{GuoLiLuo2024}} & \cref{sec:rot-ring}\\
\tcell{3.3cm}{Repulsion, quartic ring} & \tcell{6.5cm}{center empties at $\Omega_c=(3g/\pi)^{1/6}$} & \tcell{4.3cm}{TF; pure-quartic case of Refs.~\cite{FetterJacksonStringari2005,KasamatsuTsubotaUeda2002}} & \cref{sec:ring}\\
\tcell{3.3cm}{Repulsion, rotating} & \tcell{6.5cm}{$E-E_{\rm TF}[|\psi|^2]=(\epsgrad+\gflow)E_{\rm int}$} & \tcell{4.3cm}{identity} & \cref{sec:rot-rep}\\
\tcell{3.3cm}{Endpoint, $s=-2$} & \tcell{6.5cm}{$E_{\eps,R}=\eps^{-2}e_{R/\eps}(g)$; $|\psi|^2\simeq(1+\kappa)/(2gr^2)$,   $\ln L_h\simeq g/[\pi(1+\kappa)]$} & \tcell{4.3cm}{new} & \cref{sec:endpoint}\\
\tcell{3.3cm}{Split-step imaginary time} & \tcell{6.5cm}{first-order stationary bias,   $\Delta\ell/\ell\simeq\Delta\tau/[2(s+2)\ell^2\delta]$} & \tcell{4.3cm}{new near collapse} & \cref{sec:dtau-num}\\
\end{tabular}
\end{ruledtabular}
\end{table*}

\section{Nonrotating condensates}
\label{sec:conc}

\subsection{Attraction: threshold and concentration}
\label{sec:conc-att}
The two-dimensional scan covers $\delta$ from $0.4$ to $0.00625$ for $s=1,2,3,4$ and from
$0.8$ to $0.05$ for $s=-0.5,-1,-1.5$ at $\eps=0.03$ and $0.06$; points whose trial width
is below five grid spacings are not run. The radial solver extends the confining
scan to $\delta\simeq7\times10^{-4}$ and adds the unregularized singular traps,
$\eps=0$, down to $\delta=10^{-3}$. The width of a computed state is
$\ell=(2E^{\rm int}_{\rm kin})^{-1/2}$, where $E^{\rm int}_{\rm kin}$ is the kinetic energy
minus that of the center-of-mass motion (\cref{app:numerics}); for $Q_\ell$ this is
exactly $\ell$.

\emph{Threshold and trial bound.} The threshold is fixed by \cref{eq:astar}; the
computations only check consistency with it. On the supercritical side,
$\delta=-0.02$ and $-0.05$, all 20 two-dimensional runs concentrate to the grid scale: 16
fall below the flag $\ell<2.5\Delta x$ of \cref{app:numerics}, and the other four end at
$\ell\le2.9\Delta x$ when their run limit is reached. On the subcritical side the radial
energies satisfy
$E_{\rm rad}\le\Evar$ [\cref{eq:Erad}] at all 85 sampled points, with a relative gap of at least
$9\times10^{-5}$, at least $3.5$ times the radial uncertainty. This tests the radial
solver and measures how close the Townes trial comes to the ground state.

\begin{figure*}[t]
  \centering
  \includegraphics[width=\textwidth]{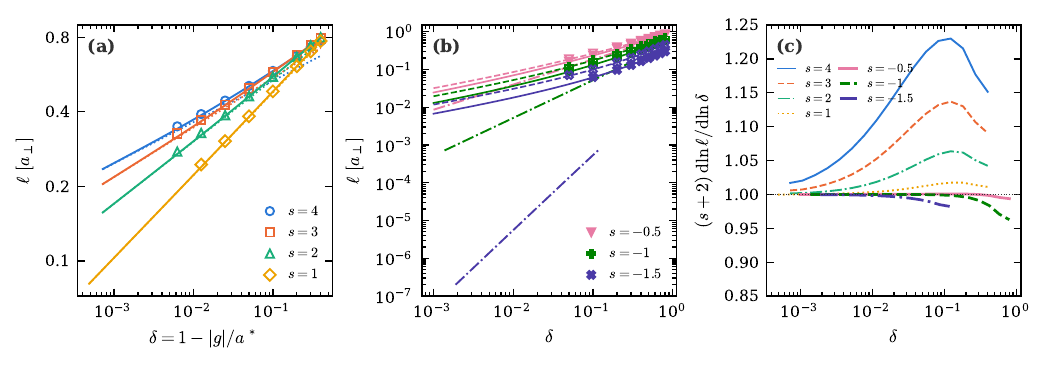}
  \caption{Near-threshold concentration, $\Omega=0$. (a) Confining traps: width $\ell$
    vs $\delta=1-|g|/g_{*}$ from the two-dimensional solver (symbols) and the
    converged radial solutions (solid), with the law \cref{eq:conc} (dotted).
    (b) Singular traps: two-dimensional states at $\eps=0.03$ (filled) and $0.06$
    (open), radial solutions at the same regulators (solid, dashed), and
    unregularized radial solutions (dash-dotted). (c) Local exponent
    $(s+2)\,\dd\ln\ell/\dd\ln\delta$ of the unregularized radial solutions for all seven
    exponents.}
  \label{fig:conc}
\end{figure*}

\emph{Concentration law.} \Cref{fig:conc} shows the result. In the confining family
[\cref{fig:conc}(a)] the local exponent $\dd\ln\ell/\dd\ln\delta$ of the radial solutions
is $0.170$, $0.2015$, $0.2505$, and $0.3334$ at $\delta=10^{-3}$ for $s=4,3,2,1$, against
$1/(s+2)=0.167$, $0.200$, $0.250$, and $0.333$; at $\delta\simeq7\times10^{-4}$,
$\ell/\ell_\ast=1.0001$--$1.0026$, so the law \cref{eq:conc} is reached from above, more
slowly for larger $s$. The Townes trial with the actual trap is closer still: at
$\delta=0.00625$, $\ell/\lvar=1.0007$--$1.017$, and the energy gap $(\Evar-E)/|\Evar|$
closes from $1.4$--$18\%$ at $\delta=0.4$ to $0.04$--$1.5\%$ at $\delta=0.00625$.

In the singular family [\cref{fig:conc}(b)] the unregularized radial widths obey the
same law with the diverging exponents $1/(s+2)=2/3$, $1$, and $2$: the local
exponents are $0.6665$ and $0.9998$ near $\delta=10^{-3}$ for $s=-0.5$ and $-1$, and
$1.9990$ for $s=-1.5$ over the last accepted interval $\delta=1.9$--$2.6\times10^{-3}$,
where the width is already $2.0\times10^{-7}\aho$. \Cref{fig:conc}(c) collects the
local exponents of all seven unregularized families: $(s+2)\,\dd\ln\ell/\dd\ln\delta\to1$.

\subsection{Regularized singular cores}
\label{sec:cone-num}
At fixed $\eps$ the widths depart from the singular law
once $\ell_\ast$ approaches $\eps$, as \cref{eq:cone} predicts. At $\eps=0.03$ the radial
widths agree with the regularized Townes trial to better than $6\times10^{-5}$
(relative) down to $\delta=10^{-4}$ for $s=-1$ and $3\times10^{-4}$ for $s=-1.5$, with local
exponents $0.38$ and $0.39$ over the smallest intervals, i.e., in the crossover. Along
the trial the local exponent decreases to $0.3345$, and the ratio of $\ell$ to the cone
width $\ell_\ast$ of \cref{eq:cone} to $1.003$, at $\delta=10^{-9}$, where
$\ell\simeq\eps/300$. The width is thus not limited by the
regulator; it crosses over to the cone law. The threshold, by contrast, is the
same with and without regularization, which is why it has to be identified through
the energy functional and not through the peak density.

\subsection{Repulsion: Thomas--Fermi scaling}
\label{sec:tf-num}
\begin{figure}[t]
  \centering
  \includegraphics[width=\columnwidth]{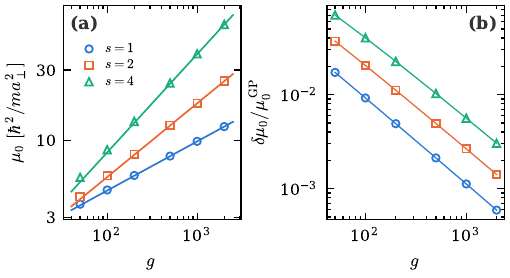}
  \caption{Repulsive gases, $\Omega=0$, $s=1,2,4$. (a) GP chemical potential of the
    converged radial states (symbols) and the parameter-free TF value \cref{eq:RTF}
    (lines). (b) Relative TF error $\delta\mu_0/\mu^{\rm GP}_0$.}
  \label{fig:mu}
\end{figure}

For $g>0$ the dilation balance sets the TF scales of \cref{eq:RTF}. \Cref{fig:mu} compares
the zero-rotation GP chemical potential $\mu^{\rm GP}_0$ of the converged radial states with
the parameter-free \cref{eq:RTF} for $s=1,2,4$. The relative error
$\delta\mu_0/\mu^{\rm GP}_0=|\mu^{\rm TF}_0-\mu^{\rm GP}_0|/\mu^{\rm GP}_0$ decreases
monotonically to $5.9\times10^{-4}$, $1.4\times10^{-3}$, and $3.0\times10^{-3}$ at $g=2000$, with
descriptive log--log slopes $-0.91$, $-0.89$, and $-0.85$ over $g=50$--$2000$; we do not
claim an asymptotic correction law. Resolving these relative errors requires
continuum-converged values; the radial uncertainty is below $1.3\times10^{-6}$ (relative). In
the singular traps the repulsive cloud is held by the box, and its TF density
$(\mu^{\rm TF}-V_{s,\eps})_+/g$ carries a central cusp; that case is tested together with
rotation in \cref{sec:rot-rep}.

\subsection{Accuracy of the two-dimensional states and the stationary bias}
\label{sec:dtau-num}
Against the
radial references the two-dimensional energies agree to within $2.5\times10^{-3}$
(confining) and $1.1\%$ (singular, $\eps=0.03$, $\Delta x=0.39\,\eps$), and the widths to
within $3.5\%$ and $4.5\%$; at $\delta=0.4$ the differences are below $2\times10^{-6}$ and
$4\times10^{-4}$. In the singular traps the two-dimensional energy lies below the
continuum value, a grid effect of the sampled regularized core. Near threshold in
the confining traps it lies above, the states are too wide, and grid doubling
changes neither by more than $10^{-6}$: the offset is a time-step effect.

The fixed point of \cref{eq:split} is not the ground state, as is known for discretized
normalized gradient flows in general~\cite{Bao2004,ChenDongLiuXie2023,WangCaiSongWang2026};
what matters here is the size of the effect near a collapse threshold. Expanding the step
to first order in $\Delta\tau$, the second half step sees the density
$|\psi_2|^2=|\psi|^2[1-\Delta\tau(2\mu-U)]+O(\Delta\tau^2)$ of the intermediate state $\psi_2$
that enters it, and the fixed point is the ground state of the modified operator
\begin{equation}
  H_{\Delta\tau}=H_\Omega-\frac{\Delta\tau}{2}\,g\,(2\mu-U)\,|\psi|^2+O(\Delta\tau^2)
  \label{eq:Hdt}
\end{equation}
(\cref{app:dtau}). The correction is first order; it is repulsive in the core of an
attractive near-threshold state, so the fixed point is too wide. Solving
\cref{eq:Hdt} self-consistently at $s=4$, $\delta=0.1$ reproduces the width shift of the
production step to $0.9\%$, $0.4\%$, $0.2\%$, and $0.02\%$ for
$\Delta\tau=2,1,0.5,0.25\times10^{-3}$ (\cref{tab:dtau}). Near threshold the shift is
amplified along the breathing mode: treated as a shift of $\delta$, the bias is
$\Delta\delta\approx0.5\,\Delta\tau/\ell^2$ at both $\delta=0.1$ and $0.00625$, and since
$\Delta\ell/\ell\simeq(\dd\ln\ell/\dd\ln\delta)\,\Delta\delta/\delta\to\Delta\delta/[(s+2)\delta]$, its effect on
the width grows as $1/\delta$.
At $\delta=0.00625$ the production step gives $\ell=0.35600$, $0.34942$, and $0.34570$ for
$\Delta\tau=4$, $2$, and $1\times10^{-4}$, against the radial $0.34162$: a $2.3\%$ error at
the step of the scan, with a residual $\etaGP=1.6\times10^{-3}$ that meets the gate. Along
a nearly singular direction a small residual is compatible with a width error of
several percent, and short $\Delta\tau/2$ checks do not relax the soft mode. Two remedies
work: extrapolation $\Delta\tau\to0$, which gives $\ell=0.34170$, and a symmetric step that
evaluates $U$ once on the normalized state and uses it in both half steps. The latter
has a second-order stationary bias: $\ell=0.341528$ and $0.341594$ at $\Delta\tau=4$ and
$2\times10^{-4}$, a bias of $2.6\times10^{-4}$ instead of $4.2\%$ at the same step. Re-evaluating
the density after renormalizing the intermediate state does not help
(\cref{tab:dtau}). These statements concern the stationary state of the map, not the
accuracy of the nonlinear flow. Local exponents from the two-dimensional widths alone are
biased by up to $0.02$; the values quoted above are radial.

\begin{table}[t]
\caption{Stationary width $\ell$ of the imaginary-time map at $s=4$, $\Omega=0$ for the
production step \cref{eq:split} (A), a step that renormalizes the intermediate state
before computing $U'$ (B), and the symmetric step with $U'=U$ (C), together with the
first-order prediction from \cref{eq:Hdt}. Radial references: $\ell=0.583727$
($\delta=0.1$) and $0.341616$ ($\delta=0.00625$). Replica of the solver step on $128^2$
points ($\delta=0.1$, $\Delta x=0.0625$); at $\delta=0.00625$ column A is from the production
runs and column C from the replica, both with $\Delta x=0.0234$.}
\label{tab:dtau}
\begin{ruledtabular}
\begin{tabular}{ccccc}
$\Delta\tau$ & A & B & C & \cref{eq:Hdt}\\ \hline
\multicolumn{5}{c}{$\delta=0.1$}\\
$2\times10^{-3}$ & 0.587014 & 0.586605 & 0.583700 & 0.587044\\
$1\times10^{-3}$ & 0.585396 & 0.585187 & 0.583720 & 0.585403\\
$5\times10^{-4}$ & 0.584568 & 0.584462 & 0.583726 & 0.584570\\
$2.5\times10^{-4}$ & 0.584149 & 0.584096 & 0.583727 & 0.584149\\ \hline
\multicolumn{5}{c}{$\delta=0.00625$}\\
$4\times10^{-4}$ & 0.356002 & --- & 0.341528 & ---\\
$2\times10^{-4}$ & 0.349418 & --- & 0.341594 & ---\\
$1\times10^{-4}$ & 0.345702 & --- & --- & ---\\
\end{tabular}
\end{ruledtabular}
\end{table}

\section{Rotating condensates}
\label{sec:rot}

Rotation acts differently on the two signs of $g$. An attractive cloud carries angular
momentum by moving its center; a repulsive one by vortices. Both are controlled by the
dilation balance: the first through the dipole frequency of the concentrating cloud, the
second through the TF size and the reduced rotation $\zeta$.

\subsection{Attraction: leaving the axis}
\label{sec:rot-axis}
\emph{Stiff dipole: rotation-independent states.} For $s=2$, $\Omega=0.9$ and for the
boxed $s=-1$, $\Omega=0.5$, the lowest states found coincide with the nonrotating ones:
$|\Delta E|\le10^{-8}$ and $\avg{\Lz}=0$ at every sampled $\delta$. This is the numerical
counterpart of Refs.~\cite{LNR2018,GuoLuoPeng2023,DinhNguyenRougerie,LeNguyen2026} and of the
dipole criterion: in the harmonic trap $\omega_1=1>\Omega$ (Kohn), and in the singular traps
the dipole stiffens near threshold.

\begin{figure}[t]
  \centering
  \includegraphics[width=\columnwidth]{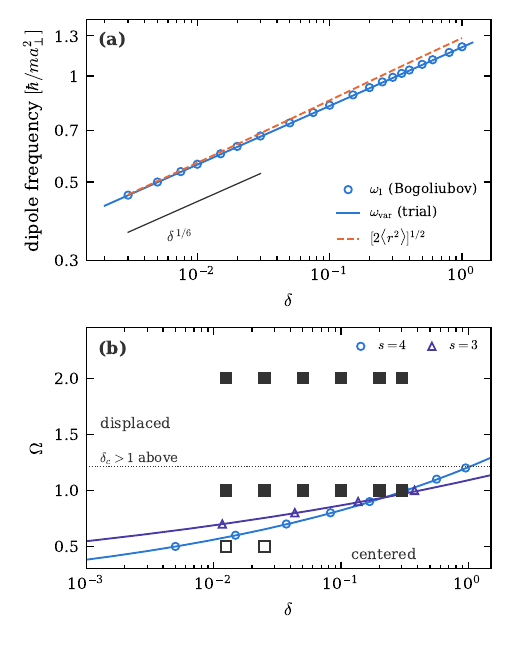}
  \caption{Leaving the axis, quartic trap. (a) Dipole frequency of the centered state:
    Bogoliubov $\omega_1$ (symbols), trial $\omega_{\rm var}=(2m_2)^{1/2}\ell_\ast$ (solid), and
    sum-rule bound $[2\avg{r^2}]^{1/2}$ on the GP state (dashed); the guide shows
    $\delta^{1/6}$. (b) Stability diagram: the centered state is unstable above the curves,
    i.e., for $\delta$ below the GP threshold $\dGP(\Omega)$ (symbols, $s=4$ circles, $s=3$
    triangles) or the closed form $\dvar$ (lines). Large markers: lowest two-dimensional
    states found at $s=4$, centered (open) or displaced (filled).}
  \label{fig:dipole}
\end{figure}

\emph{Soft dipole: the threshold.} For the quartic trap \cref{fig:dipole}(a) shows the
Bogoliubov dipole frequency of the centered state. It follows the trial value
\cref{eq:wvar}, $\omega_{\rm var}=(2m_2)^{1/2}\ell_\ast$, to within $0.14$--$0.5\%$ for
$\delta=0.003$--$1$ and softens as $\delta^{1/6}$; the sum-rule estimate
$\omega_1^2\le\avg{\partial_x^2V}=2\avg{r^2}$, evaluated on the GP state, exceeds it by
$0.4$--$6\%$. The higher sectors are stiffer: $\omega_2/2$ and $\omega_3/3$ exceed $\omega_1$
by factors between $1.079$ (at $\delta=1$) and $9.7$ (at $\delta=0.003$), and for $s=3$ by at
least $1.049$ and $1.089$ over the same range (\cref{app:bogo}), so the dipole sets the
threshold. Setting $\omega_1(\dGP)=\Omega$ gives the GP threshold $\dGP(\Omega)$ of
\cref{tab:dc}. It lies $1$--$3\%$ below the closed form \cref{eq:dcvar} for $s=4$ and
$\Omega=0.5$--$1.2$, and $2$--$10\%$ below it for $s=3$, where the trial is less accurate
at the larger $\delta$ involved. The $\Omega^6$ law of the quartic threshold is therefore
a property of the GP ground state, not only of the trial. For $\Omega\gtrsim1.21$ the
threshold moves to $\delta>1$, i.e., to repulsive coupling: the centered state is
unstable for every attractive $g$.

\begin{table}[t]
\caption{Threshold $\dGP(\Omega)$ at which the centered state becomes unstable,
from $\omega_1(\dGP)=\Omega$ with the Bogoliubov frequency $\omega_1$, compared with the
closed form $\dvar$ of \cref{eq:dcvar}. Uncertainty of $\dGP$ below $10^{-4}$
(relative).}
\label{tab:dc}
\begin{ruledtabular}
\begin{tabular}{ccccc}
$s$ & $\Omega$ & $\dGP$ & $\dvar$ & $\dGP/\dvar$\\ \hline
4 & 0.5 & 0.005003 & 0.005062 & 0.988\\
4 & 0.6 & 0.01482 & 0.01511 & 0.980\\
4 & 0.7 & 0.03714 & 0.03811 & 0.975\\
4 & 0.8 & 0.08256 & 0.08492 & 0.972\\
4 & 0.9 & 0.1675 & 0.1722 & 0.973\\
4 & 1.0 & 0.3162 & 0.3239 & 0.976\\
4 & 1.1 & 0.5622 & 0.5739 & 0.980\\
4 & 1.2 & 0.9505 & 0.9673 & 0.983\\ \hline
3 & 0.7 & 0.01167 & 0.01188 & 0.983\\
3 & 0.8 & 0.04330 & 0.04514 & 0.959\\
3 & 0.9 & 0.1361 & 0.1466 & 0.928\\
3 & 1.0 & 0.3767 & 0.4204 & 0.896\\
\end{tabular}
\end{ruledtabular}
\end{table}

\subsection{Attraction: across the threshold and on the ring}
\label{sec:rot-ring}
\begin{figure*}[t]
  \centering
  \includegraphics[width=0.94\textwidth]{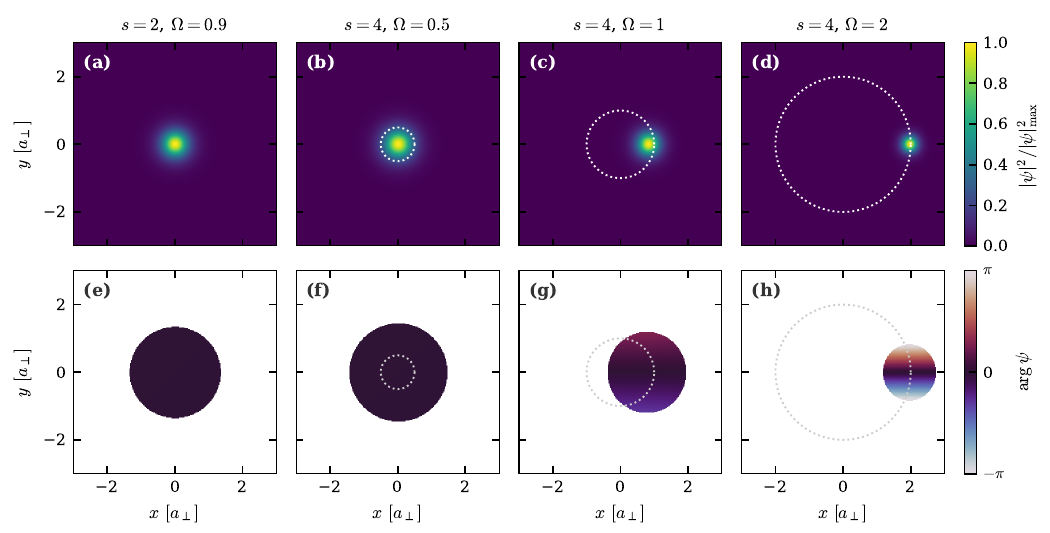}
  \caption{Lowest rotating attractive states found at $\delta=0.05$ (states of the
    original runs, not extrapolated in $\Delta\tau$; widths differ by $\le1.5\%$). Top:
    density; bottom: phase, masked below $2\times10^{-3}$ of the peak. (a,e) $s=2$,
    $\Omega=0.9$: centered, vortex free, identical to $\Omega=0$. (b,f) $s=4$, $\Omega=0.5$:
    centered. (c,g) $\Omega=1$ and (d,h) $\Omega=2$: displaced lump approaching the
    minimum ring of $\Veff$ (dotted circle of radius $r_0=\Omega$); its phase is the boost
    phase $\Omega x_0y$, with $y$ the laboratory coordinate and the lump on the $x$ axis.}
  \label{fig:states}
\end{figure*}

\begin{figure*}[t]
  \centering
  \includegraphics[width=\textwidth]{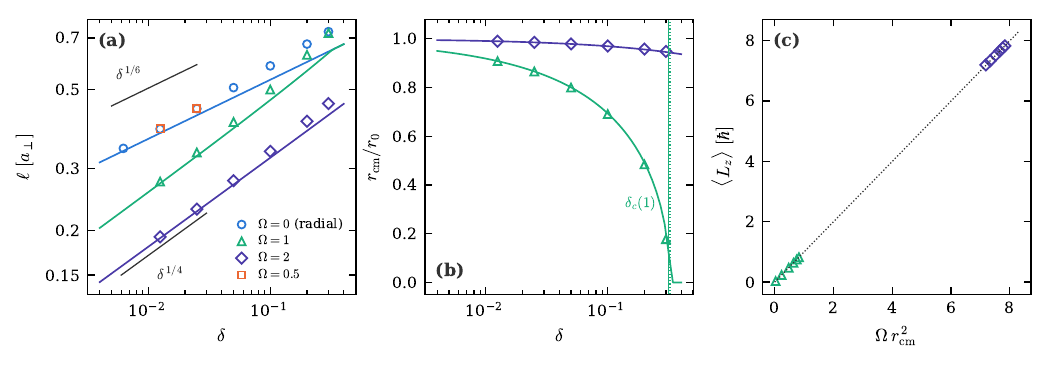}
  \caption{Rotating quartic trap, lowest states found, extrapolated to $\Delta\tau\to0$
    (symbols), and the Townes trial \cref{eq:quartic} (lines). (a) Width vs $\delta$ at
    $\Omega=0$ (converged radial states), $0.5$ (identical to $\Omega=0$), $1$, and $2$, with
    guides $\delta^{1/6}$ and $\delta^{1/4}$. (b) Center of mass $r_{\rm cm}/r_0$; dotted:
    trial threshold $\dvar(1)=0.324$; solid: GP threshold $\dGP(1)=0.316$. (c) $\avg{\Lz}$
    vs $\Omega r_{\rm cm}^2$ (dotted: equality).}
  \label{fig:rot}
\end{figure*}

\Cref{fig:states,fig:rot} show the
rotating quartic trap at $\Omega=0.5$, $1$, and $2$, for which $r_0=\Omega$. Because the
widths are near-threshold quantities, the lowest states were recomputed at
$\Delta\tau=8$, $4$, $2$, and $1\times10^{-4}$ and extrapolated to $\Delta\tau\to0$ from the three
smallest steps (\cref{app:ring}). On four nonrotating controls this protocol
reproduces the radial widths to $2\times10^{-6}$--$1.3\times10^{-3}$ and the energies to
$1.2\times10^{-4}$. Several stages, including all stages of the displaced states at
$\Omega=1$, reached their time limit before the stopping criterion (\cref{app:ring}); the
estimated remaining drift of their widths is at most $3.4\times10^{-5}$ except at
$\delta=0.3$ (\cref{tab:ring}). The difference between extrapolations from the three largest
and the three smallest steps, which we take as the uncertainty, is at most
$2.2\times10^{-3}$ in $\ell$ and $1.7\times10^{-4}$ in $E$ (relative) for the rotating states,
and below $1.5\times10^{-4}$ in $\ell$ for $\delta\ge0.025$. Refining the grid spacing from $0.023$ to
$0.0125$ at fixed $\Delta\tau$ changes $\ell$, $r_{\rm cm}$, and $E$ of the $\Omega=2$,
$\delta=0.05$ lump by less than $3\times10^{-8}$ (relative; \cref{app:ring}).

The states agree with the threshold of \cref{tab:dc} [\cref{fig:dipole}(b)]. At
$\Omega=0.5$ all sampled points lie above $\dGP=0.0050$, and the lowest state is
centered and identical to the nonrotating one ($|\Delta E|<10^{-12}$). At $\Omega=2$ the
threshold lies at repulsive coupling, and every lowest state is a lump near the ring,
$r_{\rm cm}/r_0=0.95$--$0.99$, with $r_{\rm cm}$ the distance of the center of mass from
the axis. At $\Omega=1$, $\dGP=0.3162$ lies just above the largest sampled $\delta=0.3$.
There the centered state is a stationary point with $\omega_1=0.9913<\Omega$, hence not a
minimum, and the displaced state found lies $1.4\times10^{-4}$ below it. Its displacement,
$r_{\rm cm}=0.177$ (a time-limited run), satisfies $r_{\rm cm}^2/(\dGP-\delta)=1.94$, close
to the trial slope $1.84$ of $x_0^2/(\dvar-\delta)$, as expected for a continuous
transition with the same quartic coefficient. As $\delta$ decreases the lump moves out:
$r_{\rm cm}/r_0=0.484$, $0.691$, $0.799$, $0.865$, and $0.908$ for $\delta=0.2$, $0.1$, $0.05$,
$0.025$, and $0.0125$, against the trial $0.491$, $0.693$, $0.800$, $0.865$, and $0.908$.

On the ring the width approaches \cref{eq:ring} from above. At $\Omega=2$,
$\ell/\ell_\ast=1.010$, $1.017$, and $1.028$ at $\delta=0.0125$, $0.025$, and $0.05$, and the
two-point exponent over $\delta=0.0125$--$0.025$ is $0.260$, against $1/4$; the nonrotating
state over the same interval gives $0.190$, against $1/6$. The center tracks $x_{0,\rm
var}$ to $0.3\%$, and the energy lies below the trial bound (by $1.9\times10^{-3}$ at
$\delta=0.0125$). At $\Omega=1$ the width converges more slowly, $\ell/\ell_\ast=1.023$ at
$\delta=0.0125$, because the center has not yet reached the ring. In all displaced states
$\avg{\Lz}=\Omega r_{\rm cm}^2$ to within $0.5\%$ [\cref{fig:rot}(c)]: the angular momentum
is that of rigid center-of-mass rotation~\cite{WilkinGunnSmith1998,Lundh2004,Collin2005},
and the internal structure feels $\nabla^2V$, as \cref{sec:ring} predicts.

\emph{Prepared vortices.} Charge-$q$ ring states with $q=1,2,3$ at $s=4$, $\Omega=2$,
$|g|/g_{*}=0.5$ and $0.9$, perturbed by $1\%$ azimuthal noise and relaxed in imaginary time
for $\tau=25$ ($q=1,3$) or $40$ ($q=2$), all end on the displaced lump: their energies,
$-2.061676$ and $-3.061154$, agree with the lump to $10^{-6}$, $r_{\rm cm}=1.861$ and
$1.940$, and the central winding vanishes. Imaginary-time relaxation thus selects the
vortex-free lump, in line with the vortex-free ground states of
Refs.~\cite{GuoLuoPeng2023,GuoLiLuo2024,LeNguyen2026}; it does not test the real-time
stability of a vortex.

\subsection{Repulsion: vortex lattices and the Thomas--Fermi error}
\label{sec:rot-rep}
For $g>0$ rotation nucleates vortices instead of moving the cloud. The total kinetic energy
is then dominated by the rigid part of the circulation, which the rotating-frame TF density
already contains, and the TF energy error is the sum $\epsgrad+\gflow$ of
\cref{eq:epsgrad}. For a harmonic vortex lattice with $21$ vortices at $g=1000$ and
$\Omega=0.7$ (\cref{app:rep}) the total kinetic-to-interaction ratio is $0.966$, and rigid
rotation alone, $\half\Omega^2\avg{r^2}/E_{\rm int}$, gives $0.960$; the two parts omitted by
TF are $\epsgrad=0.057$ and $\gflow=0.138$. In a lattice the flow part, the mismatch between
the vortex flow and rigid rotation, is the larger part of the TF error. In the window
$\Omega\in[0.3,0.9]$ for $s=2$ and $[0.5,2]$ for $s=4$, between the rigorous slow- and
fast-rotation regimes, $\epsgrad$ decreases with $g$ at every sampled point; its largest
value, $0.192$, occurs at $s=2$, $g=500$, $\Omega=0.9$ (\cref{tab:epsgrad}). All quartic points
lie below the annular threshold, $\zeta<\zeta_c$.

The singular traps test these diagnostics where TF has a cusp. In the boxed traps
($\Rbox=7$, $\eps=0.02$, $512^2$ points on a square of side $20$, so that $\eps/\Delta x=0.51$)
we compare the azimuthally averaged GP density with the independently normalized
rotating-frame TF density through the relative radial discrepancy $\delta n_{2,r}$ of
\cref{eq:dn}. The grid does not resolve the regularized core, and the residual shows when
this matters. At $\Omega=0$, where converged radial states exist for the same traps, the
production grid and step reproduce $\epsgrad$ to $7\times10^{-4}$ and $\delta n_{2,r}$ to
$1.7\%$ for $s\ge-1$, with $\etaGP\le0.009$; at $s=-1.5$, $-1.75$, and $-1.9$ ($g=1000$) the
errors grow to $1.7\%$, $16\%$, and $61\%$ in $\epsgrad$ and to $15\%$, $44\%$, and $54\%$
in $\delta n_{2,r}$, together with $\etaGP=0.034$, $0.15$, and $0.43$ (\cref{tab:cal}). The
errors come from the grid: at $s=-1.75$, quartering $\Delta\tau$ lowers $\etaGP$ to $0.014$
but leaves them at $14\%$ and $45\%$, whereas $1024^2$ points reduce them to $2.3\%$ and
$14\%$. On the production grid the residual grows with the grid error, because both grow
with the depth of the cusp, and every calibration state with $\etaGP\le0.02$ has errors below
$0.7\%$ in $\epsgrad$ and $9\%$ in $\delta n_{2,r}$; refining a gated rotating state to
$1024^2$ points changes the two by $0.5\%$ and $6\%$ (\cref{app:rep}). We therefore draw
conclusions only from states that meet the gate $\etaGP\le0.02$. At
$\Omega=0.3$, 14 of the 30 computed states do; they cover $-1.5\le s\le-0.25$ and
$g=500$--$2000$. For all of them $\gflow=(0.81$--$0.89)\,\epsgrad$: the flow part of the TF
error is nearly as large as the density-gradient part. Both decrease with $g$, as does
$\delta n_{2,r}$; at $s=-1$, $\delta n_{2,r}=0.334$, $0.212$, and $0.140$ for $g=500$, $1000$,
and $2000$ [\cref{fig:reprot}(a)]. The Spearman rank correlation between $\epsgrad$ and
$\delta n_{2,r}$ is $0.952$, and $0.627$ after the rank of $g$ is controlled for; for the $L^1$
and radius-based discrepancies the controlled coefficients lie between $-0.93$ and $-0.02$
(\cref{tab:corr}). In converged radial states at $\Omega=0$, by contrast, $\epsgrad$,
$\delta n_{2,r}$, and the $L^1$ discrepancy have the same rank order within each $g$, the error
of the rms radius nearly so (within-$g$ coefficients $\ge0.79$), and only the $90\%$ radius
departs from it. The density-gradient ratio
thus tracks the $L^2$ profile discrepancy in both settings, but in the rotating states it
does not track errors of the cloud radius.
No significance level is assigned to these designed sets.

All these states belong to the branch selected by the imprinted lattice
(\cref{sec:methods}). At $s=-1.5$, $g=2000$ a start with seven vortices ends in a stationary
state with $Q_\phi=7$ and $\avg{\Lz}=5.63$ instead of $7.53$, whose energy is lower by
$2.4\%$; its $\epsgrad$ and $\gflow$ are smaller by $13\%$ and $17\%$, so that
$\gflow/\epsgrad=0.83$, while $\delta n_{2,r}$ is larger by $7\%$ (\cref{app:rep}). The ratio
$\gflow/\epsgrad$ is thus insensitive to the vortex configuration, the individual
diagnostics change by $7$--$17\%$ between the two branches, and the rank correlations above
describe one branch.

\subsection{Repulsion in singular traps: circulation and core counting}
\label{sec:rot-circ}
\begin{figure*}[t]
  \centering
  \includegraphics[width=\textwidth]{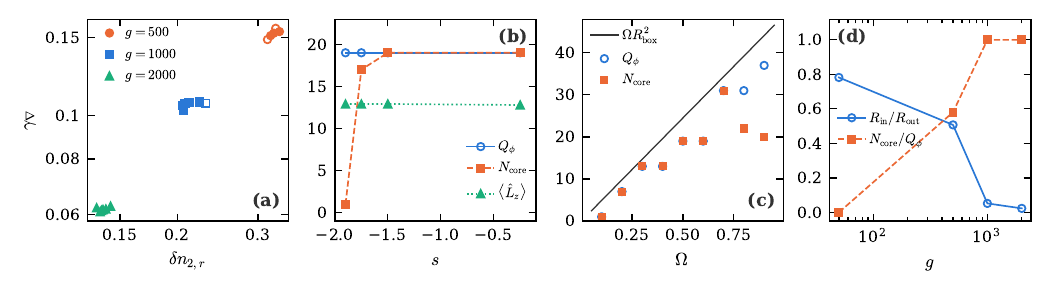}
  \caption{Rotating repulsive condensates in boxed singular traps ($\Rbox=7$, $\eps=0.02$).
    (a) Density-gradient ratio $\epsgrad$ vs radial TF discrepancy $\delta n_{2,r}$ at
    $\Omega=0.3$: filled symbols meet the residual gate $\etaGP\le0.02$, open symbols
    $0.02<\etaGP\le0.05$; the 13 states with larger residuals, all at $s\le-1.5$, are grid
    dominated (\cref{tab:cal}) and not shown. (b) Trap shape at $g=2000$, $\Omega=0.5$: net winding $Q_\phi$,
    density-resolved core count $\Ncore$, and $\avg{\Lz}$. (c) Rotation at $s=-1$, $g=1000$:
    $Q_\phi$, $\Ncore$, and the Feynman estimate $\Omega\Rbox^2$. All states belong to the branch
    selected by the imprinted lattice (\cref{sec:methods}). (d) Interaction strength at
    $s=-1$, $\Omega=0.7$, fixed $Q_\phi=31$: ratio $R_{\rm in}/R_{\rm out}$ of the inner and
    outer $10\%$ density radii, and $\Ncore/Q_\phi$.}
  \label{fig:reprot}
\end{figure*}

The singular trap shapes the vortex state through the density it imposes.
\Cref{fig:reprot}(b) follows the trap shape at $g=2000$, $\Omega=0.5$. The net phase winding
$Q_\phi$ [\cref{eq:Qphi}] and $\avg{\Lz}$ stay at $19$ and $12.79$--$12.91$ from $s=-0.25$ to
$-1.9$, and $\epsgrad$ grows only from $0.074$ to $0.088$, whereas the number $\Ncore$ of cores
counted with a density threshold referenced to the peak falls from $19$ to $17$ at $s=-1.75$
and to $1$ at $s=-1.9$. The cusp raises the peak density until the threshold exceeds the
density next to the cores; the phase winding on a contour of nonvanishing density, an
integer invariant along the branch (\cref{sec:ring}), is insensitive to this. The failure
belongs to the counting criterion, not to the vortices: near a cusp the circulation is to be
read from the phase or from $\avg{\Lz}$. The state at $s=-1.9$ meets only the near-endpoint
gate ($\etaGP=0.047$) and, by the $\Omega=0$ calibration above, its density is grid
dominated; we use it only for its winding.

Along the imprinted branch, rotation and interaction strength redistribute the
circulation in the box. At $s=-1$, $g=1000$ [\cref{fig:reprot}(c)], increasing $\Omega$
from $0.1$ to $0.9$ gives $Q_\phi=1,7,13,13,19,19,31,31,37$, below the Feynman estimate
$\Omega\Rbox^2$ of a uniformly filled box~\cite{Feynman1955}. These windings depend on the
initial lattice, and a start with fewer vortices can end lower in energy
(\cref{sec:rot-rep}), so they are not equilibrium vortex numbers. $\Ncore$ follows $Q_\phi$
up to $\Omega=0.7$ and falls below it
at $\Omega=0.8$ and $0.9$; at $\Omega=0.9$ a central hole of radius $R_{\rm in}=3.62$ opens
(the state at $\Omega=0.8$ misses the residual gate, $\etaGP=0.039$). At fixed $Q_\phi=31$ ($s=-1$,
$\Omega=0.7$), decreasing $g$ from $2000$ to $50$ raises the ratio $R_{\rm in}/R_{\rm out}$ of
the radii at which the azimuthally averaged density first and last exceeds $10\%$ of its
maximum from $0.02$ to $0.78$, while $\Ncore/Q_\phi$ falls from $1$ to $0$
[\cref{fig:reprot}(d)]: the lattice turns into a giant-vortex-like annulus, a finite-box
crossover rather than a sharp transition. The box itself remains dominant: at $s=-1$,
$g=1000$, $\Omega=0.5$, the outer radius stays at $R_{\rm out}/\Rbox=1.016$--$1.005$ for
$\Rbox=3.5$--$8.5$, with $Q_\phi=\Ncore=7,7,13,19,19,31$.

These states are fixed-grid values with an unresolved regulator, not extrapolated in
$\Delta\tau$; for those that meet the gate, the $\Omega=0$ calibration bounds the grid error of
$\epsgrad$ and $\delta n_{2,r}$ (\cref{tab:cal}). The TF states have no soft mode, so the
time-step amplification of \cref{sec:dtau-num} does not arise: quartering $\Delta\tau$ moves
$\epsgrad$ and $\delta n_{2,r}$ by at most $2\%$ even at $s=-1.75$ (\cref{app:rep}).

\section{The inverse-square endpoint}
\label{sec:endpoint}

We remove the regulator, $\eps=0.25\to0.025$ in two dimensions and down to $10^{-6}$ with
the radial solver, at $s=-1.5$, $-1.9$, and $-2$ with $g=-3$, $0$, $2$, $10$, $50$, in a
box $\Rbox=2$. The two-dimensional grid keeps $\eps/\Delta x\ge5$ with
$\Delta\tau\le0.2\,\eps^2$. We monitor the core mass
\begin{equation}
  \Mcore=\int_{|\x|<2\eps}|\psi(\x)|^2\,\dd\x,
  \label{eq:Mcore}
\end{equation}
the rms radius $r_{\rm rms}=[\int_{\R^2}|\x|^2|\psi(\x)|^2\,\dd\x]^{1/2}$ in units of $\eps$, and
$E\eps^2$. For $s>-2$ a bound state of finite width has $\Mcore\propto\eps^2$ once $\eps$ is
below its width. At $s=-2$, Proposition~1 implies that $E\eps^2\to e(g)$; whether
$\Mcore$ and $r_{\rm rms}/\eps$ also become $\eps$ independent is a property of the
profiles, which we test.

\begin{figure*}[t]
  \centering
  \includegraphics[width=\textwidth]{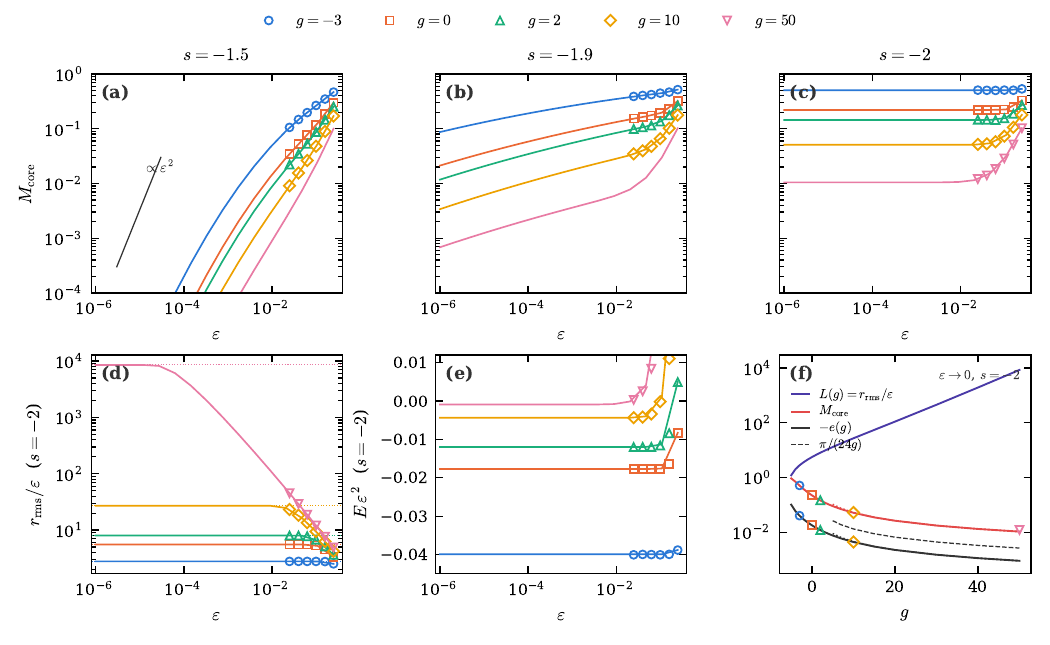}
  \caption{Regulator removal, $\Rbox=2$. Symbols: two-dimensional states; lines:
    radial solutions. (a--c) Core mass $\Mcore$ [\cref{eq:Mcore}] for $s=-1.5,-1.9,-2$.
    (d) $r_{\rm rms}/\eps$ and (e) $E\eps^2$ at $s=-2$; dotted lines in (d) are the rms
    sizes $L(g)$ of the rescaled profiles. (f) Rescaled profile [$\eps\to0$ limit of
    \cref{eq:resc}, $W=W_1$] vs $g$: $L(g)$, core mass, and $-e(g)$, with the large-$g$
    asymptotes $0.523/g$ and $0.0448/g$ (dotted) and the bound $\pi/(24g)$ (dashed);
    symbols are the two-dimensional values at $\eps=0.025$.}
  \label{fig:end}
\end{figure*}

\emph{Regulator removal.} \Cref{fig:end} shows the result. Radial and two-dimensional
data agree at every point, flagged states included:
$|\Delta(E\eps^2)|\le1.3\times10^{-4}$, $\Mcore$ to $1.1\%$, and $r_{\rm rms}$ to $0.13\%$.
At $s=-2$, for $g=-3$, $0$, and $2$, all three diagnostics are $\eps$ independent below
$\eps\simeq0.06$. The whole condensate shrinks with the regulator,
$r_{\rm rms}/\eps=2.78$, $5.57$, and $8.00$, and $E\eps^2=-0.0399$, $-0.0177$, and $-0.0119$;
these agree with the values $L(g)$ and $e(g)$ of the rescaled problem \cref{eq:resc}
solved on $D_\Lambda$, which do not change between $\Lambda=10^6$ and $10^7$ within their
uncertainties.
$\Mcore=0.506$, $0.221$, and $0.145$ is the fraction of the rescaled profile inside $u<2$,
not a partial capture.

For $g=10$ and $50$ the two-dimensional range is too short, and the radial solutions
show two crossover scales. The core converges once the box no longer compresses the
inner halo: $\Mcore$ and $E\eps^2$ are within $0.3\%$ of their limits for $\eps\lesssim0.02$
($g=10$) and $\eps\lesssim2\times10^{-3}$ ($g=50$). The whole profile, including its rms
radius, fits in the box only for $\eps\lesssim\Rbox/L_h(g)$, i.e., $\eps\lesssim0.04$ and
$\lesssim6\times10^{-5}$ [$L_h(10)=48$, $L_h(50)=3.4\times10^4$], an order-of-magnitude
criterion. The two-dimensional values at $g=50$ ($\Mcore=0.012$, $E\eps^2>0$) are
therefore box limited. Only in this box-limited regime does the soft wall matter:
where the profile fits, soft wall and Dirichlet disk give the same $E\eps^2$ to
$3\times10^{-6}$ and the same $r_{\rm rms}$ to $1.2\times10^{-3}$ (relative), whereas at $g=50$, $\eps=0.02$
the residual energy changes by a factor of $4.5$ (\cref{app:endpoint}).

\begin{figure}[t]
  \centering
  \includegraphics[width=\columnwidth]{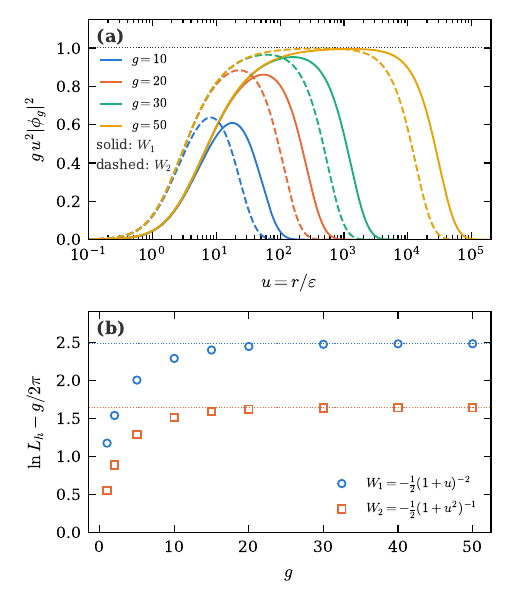}
  \caption{Rescaled endpoint profiles for two regularizations, $W_1(u)=-\frac12(1+u)^{-2}$
    (solid) and $W_2(u)=-\frac12(1+u^2)^{-1}$ (dashed). (a) Halo diagnostic
    $g\,u^2|\phi_g|^2$; the plateau at $1$ is \cref{eq:halo}. (b) $\ln L_h-g/2\pi$ with
    $L_h=(g/\pi)^{1/2}L(g)$; dotted lines: large-$g$ values.}
  \label{fig:halo}
\end{figure}

\emph{Core and halo.} The rescaled profiles carry the halo of \cref{eq:halo}
[\cref{fig:halo}(a)]: $g\,u^2|\phi_g|^2$ develops a plateau at $1$, reaching $0.994$ at $u=10^3$
for $g=50$, and its extent obeys \cref{eq:Lh} quantitatively,
$\ln L_h-g/2\pi=2.44$, $2.47$, $2.48$, and $2.48$ for $g=20$, $30$, $40$, and $50$, i.e.,
$C\simeq12$. The core saturates, $g\,\Mcore\to0.523$ and $g\,e(g)\to-0.0448$ for $g\ge20$;
the Young bound $-\pi/(24g)$ is satisfied with a factor of $2.9$ to spare. Repulsion thus
dilutes the core as $1/g$ and pushes the halo out exponentially, but does not change
the fall to the center.

\emph{What is universal.} Repeating the rescaled problem with the smooth
regularization $W_2(u)=-\frac12(1+u^2)^{-1}$ separates the features set by the tail of
the potential from those set by its core (\cref{tab:reg}, \cref{fig:halo}). The halo
plateau and the coefficient of $g$ in $\ln L_h$ are the same; $\ln L_h-g/2\pi$ saturates
at $1.64$ instead of $2.48$, i.e., $C\simeq5.1$; and the core numbers change by factors of
$2.6$--$4.1$: $g\,\Mcore\to1.386$ and $g\,e(g)\to-0.182$, within the bound $-\pi/(8g)$ that
Young's inequality gives for $W_2$. Universality of the endpoint profile therefore means
independence of $\eps$ within a fixed regularization family; the exponent of the
cutoff dependence, the halo, and the exponential growth $e^{g/2\pi}$ are independent of
the regularization as well.

\emph{Tail strength.} With $W$ multiplied by $\kappa=1/2$ and $2$ (\cref{tab:kappa}) the
plateau of $g\,u^2|\phi_g|^2$ approaches $(1+\kappa)/2$, reaching $0.734$ at $g=30$ for
$\kappa=1/2$ and $1.494$ at $g=80$ for $\kappa=2$, and $\ln L_h-g/[\pi(1+\kappa)]$ saturates at
$3.37$ and $1.91$, as \cref{eq:halokappa} predicts. The coefficient of $g$ in the exponent is
thus fixed by the tail strength alone.

\begin{table*}[t]
\caption{Rescaled endpoint problem \cref{eq:resc} for the regularizations $W_1$ and $W_2$
of \cref{fig:halo} ($\Lambda=10^7$ for $g<30$, $10^{10}$ otherwise; relative uncertainties
below $4\times10^{-6}$): energy $e(g)$, core mass $\Mcore$ ($u<2$), rms size $L(g)$, and
$\ln C(g)=\ln L_h-g/2\pi$ [\cref{eq:Lh}].}
\label{tab:reg}
\begin{ruledtabular}
\begin{tabular}{rcccc|cccc}
 & \multicolumn{4}{c|}{$W_1$} & \multicolumn{4}{c}{$W_2$}\\
$g$ & $-e$ & $\Mcore$ & $L$ & $\ln C$ & $-e$ & $\Mcore$ & $L$ & $\ln C$\\ \hline
$-3$ & 0.03993 & 0.5057 & 2.781 & --- & 0.1202 & 0.7505 & 1.800 & ---\\
0 & 0.01771 & 0.2210 & 5.565 & --- & 0.06295 & 0.4585 & 3.092 & ---\\
2 & 0.01195 & 0.1453 & 8.005 & 1.536 & 0.04487 & 0.3359 & 4.177 & 0.886\\
10 & 0.004349 & 0.05098 & 27.10 & 2.287 & 0.01750 & 0.1333 & 12.47 & 1.511\\
20 & 0.002238 & 0.02614 & 110.1 & 2.443 & 0.009073 & 0.06917 & 48.25 & 1.619\\
50 & 0.000896 & 0.01046 & 8547 & 2.479 & 0.003635 & 0.02771 & 3684 & 1.638\\
\end{tabular}
\end{ruledtabular}
\end{table*}

\emph{Neighboring exponents.} At $s=-1.5$ the radial solutions reach
$\Mcore\propto\eps^{1.9}$ below $\eps\simeq10^{-4}$, the bound-state behavior; the
two-dimensional range $\eps\ge0.025$ lies in the crossover (slopes $0.7$--$1.2$), because the
unregularized rms radii are only $0.015$--$1.1$ for $g=-3$--$50$. At $s=-1.9$ the
Townes-trial width $[(1+g/g_{*})/m_s]^{1/(s+2)}$ is below $3\times10^{-13}$ for
$-3\le g\le2$, and down to $\eps=10^{-6}$ the state behaves as a slowly drifting endpoint
profile, $\Mcore\propto\eps^{0.19}$--$\eps^{0.27}$. Near $s=-2$ the distinction between the
endpoint and its neighbors is thus asymptotic; only $s=-2$ has the exact covariance of
Proposition~1. Finally, the rotating case $\Omega=0.5$ at $s=-2$, $g=2$ is identical to
$\Omega=0$ ($\avg{\Lz}=0$): the rotation term is dilation invariant and negligible against
the $\eps^{-2}$ energies.

\section{Experimental implications and validity}
\label{sec:exp}

Box traps~\cite{Navon2021} combined with painted potentials~\cite{Gauthier2016} provide outer
confinement and radial shaping; a blue-detuned Laguerre--Gaussian ring approximates the
quartic member. The Townes regime has been reached in planar
gases~\cite{BakkaliHassani2021,ChenHung2021}, and Feshbach-tuned attractive condensates
collapse near the predicted threshold~\cite{Sackett1998,Donley2001}. With
$g=\sqrt{8\pi}Na_s/a_z$ the threshold corresponds to
$N_c=g_{*} a_z/(\sqrt{8\pi}|a_s|)\simeq1.17\,a_z/|a_s|$ atoms, independent of the trap; e.g.,
$N_c\approx1.2\times10^3$ for $a_z=1\,\mu$m and $a_s=-1$\,nm.

On the attractive side three observables follow. (i)~In an anharmonic trap with $s>2$ the
dipole (center-of-mass) oscillation softens as the atom number approaches $N_c$; for $s=4$,
$\omega_1\propto\ell\propto(N_c-N)^{1/6}$ [\cref{fig:dipole}(a)], in contrast to the
$N$-independent Kohn frequency of a harmonic trap. (ii)~In a trap stirred at $\Omega$, the
collapse moves off the axis once $\omega_1<\Omega$, i.e., for $N$ above the value that
corresponds to $\dGP(\Omega)$; the displaced cloud orbits rigidly with angular momentum
$\Omega r_{\rm cm}^2$ per atom, visible in time of flight. (iii)~The in-situ width scales as
$(N_c-N)^{1/(s+2)}$ while $N_c$ does not depend on the trap; a smooth or conical core changes
the exponent to $1/4$ or $1/3$. On the repulsive side, (iv)~the TF radius grows as
$N^{1/(s+2)}$ with the same exponent, and in a stirred quartic trap the center empties at the
reduced rotation $\zeta_c=(3/\pi)^{1/6}$, i.e., at $\Omega_c\propto N^{1/6}$; (v)~near a
singular core, counts of vortex cores resolved in the density underestimate the net phase
winding, so that the circulation is better read from the phase or from $\avg{\Lz}$. The
dipole criterion is energetic: in an isotropic trap the cloud reaches the displaced state
through dissipation by the thermal component or through the anisotropy of the stirring
potential. The endpoint predictions require a potential that follows $-\kappa/(2r^2)$ over a
range of radii. Polarizable atoms around a charged wire realize such a tail in two
dimensions~\cite{Denschlag1998}: the line charge sets $\kappa$, and the wire radius plays the
role of $\eps$. A focused Gaussian dimple has no inverse-square tail. In an inverse-square
potential the halo density $N|\psi|^2\simeq(1+\kappa)a_z/(2\sqrt{8\pi}a_s r^2)$ does not
depend on $N$, and its extent $\eps L_h(g)$ grows exponentially with $N$ until it reaches the
end of the inverse-square range.

The mean-field description has four limits here. First, the two-dimensional GP functional
requires every length to exceed $a_z$, so the covariant endpoint regime is the window
$a_z\ll\eps\lesssim\Rbox/L_h(g)$, which closes at strong repulsion. Second, for $W_1$ and
$g\gtrsim20$ the endpoint core holds $N\Mcore\simeq0.523\,a_z/(\sqrt{8\pi}\,a_s)\approx0.10\,a_z/a_s$
atoms, independent of $N$ (about $20$ for $a_z=1\,\mu$m and $a_s=5$\,nm; $2.6$ times more for
$W_2$); for $a_s\gtrsim a_z/10$ the core is a few-body object, and only the halo, which holds
most of the atoms, remains a mean-field prediction; many-body calculations in three
dimensions find a metastable gas separated from the collapsed state by a barrier that grows
with $N$~\cite{AstrakharchikMalomed2015}. Third, in a quasi-two-dimensional gas the coupling depends logarithmically on the
relevant momentum $k$~\cite{PetrovHolzmannShlyapnikov2000},
\begin{equation}
  g\to\frac{g}{1+\dfrac{a_s}{\sqrt{2\pi}\,a_z}\ln\dfrac{1}{\pi k^2a_z^2}} ,
  \label{eq:g2D}
\end{equation}
a relative correction $\sim(a_s/a_z)\ln(\ell/a_z)$ for a structure of size $1/k\sim\ell$;
it is largest for the halo, where $\ln(\eps L_h/a_z)\ge g/2\pi$ gives at least
$\simeq(2/\pi)Na_s^2/a_z^2$. The dilation symmetry itself is a mean-field
property~\cite{PitaevskiiRosch1997,Olshanii2010}. Finally, three-body loss limits the
lifetime of dense near-critical and endpoint states~\cite{Donley2001}, and quintic or
Lee--Huang--Yang terms can arrest the endpoint
collapse~\cite{NguyenRicaud2022,Shamriz2020,Malomed2018}. The calculations are at zero
temperature; thermal phase fluctuations of the quasi-two-dimensional gas, which matter most
for the dilute halo, are not included. All conclusions concern the regularized, boxed
mean-field models specified above.

\section{Conclusion}
\label{sec:conclusion}

In two dimensions a single dilation balance sets the size of a trapped condensate for both
signs of the interaction: the cloud size obeys $L^{s+2}\propto\delta$ for attraction and
$L^{s+2}\propto g$ for repulsion. On the attractive side the contact coupling fixes whether
the condensate collapses and the trap fixes only how, through its local shape where the
cloud concentrates; the law $\ell\propto\delta^{1/(s+2)}$ holds for seven exponents, and a
regularized singular core replaces it by the cone law $\delta^{1/3}$. On the repulsive side
the same exponent gives the TF radius, the trap acts through its global shape, and the
energy that TF omits splits exactly into a density-gradient part and a flow part, neither
of which counts the rigid circulation; the flow part is comparable to the density-gradient
part in the singular traps and $2.4$ times larger in a harmonic vortex lattice.

Rotation separates the two signs and then brings them to the same place. An attractive cloud
leaves the axis when $\Omega$ exceeds its dipole frequency, $\propto\ell^{(s-2)/2}$ near
threshold, at $\delta\simeq0.324\,\Omega^6$ in the quartic trap (Bogoliubov: within $3\%$),
and collapses on the ring as a rigidly rotating lump whose width is set by the bare-trap
curvature. A repulsive cloud fills with vortices, and its TF density empties at
$\zeta_c=(3/\pi)^{1/6}$ into an annulus on the same ring. For $0<s<2$ there is no ring and
only a box holds a rotating cloud; in boxed singular traps the phase winding, not a
density-based count of cores, measures the circulation near the cusp.

At the inverse-square endpoint no balance is left: the energy is exactly covariant with the
cutoff, the condensate falls to the center for either sign of $g$, and repulsion only spreads
a halo of density $(1+\kappa)/(2gr^2)$ and extent $\propto e^{g/[\pi(1+\kappa)]}$, set by the same
logarithm that prevents repulsion from arresting the fall. Near a collapse threshold, split-step imaginary
time carries a first-order stationary bias that a residual gate does not detect; a symmetric
step removes it to second order. Open questions are the real-time dynamics of off-axis collapse, the giant-vortex
onset on the quartic ring beyond the TF threshold, the finite-temperature fate of the endpoint
core, and its arrest by higher-order repulsion.


\appendix

\section{Townes-trial calculations}
\label{app:trial}

\emph{Boosted lump.} For $\psi=e^{i\mathbf p\cdot\x}Q_\ell(\x-\x_0)$ with real $Q$,
\begin{align}
  \half\|\nabla\psi\|_2^2&=\frac1{2\ell^2}+\frac{|\mathbf p|^2}{2},\qquad
  \frac g2\|\psi\|_4^4=\frac{g}{2g_{*}\ell^2},\nonumber\\
  \avg{\psi,\Lz\psi}&=(\x_0\times\mathbf p)\cdot\hat{\mathbf z}.
  \label{eq:trialterms}
\end{align}
Minimizing $|\mathbf p|^2/2-\Omega(\x_0\times\mathbf p)\cdot\hat{\mathbf z}$ gives
$\mathbf p=\Omega\hat{\mathbf z}\times\x_0$, the value $-\half\Omega^2x_0^2$, and
$\avg{\Lz}=\Omega x_0^2$; together with $\avg{V}=\avg{V(|\x_0+\ell\y|)}_Q$ this is
\cref{eq:Eoff}. In the magnetic form \cref{eq:Emag} the same trial has
\begin{align}
  \half\|(-i\nabla-\mathbf A)\psi\|_2^2&=\frac1{2\ell^2}+\frac{\Omega^2\ell^2m_2}{2},\nonumber\\
  \Big\langle-\frac{\Omega^2r^2}{2}\Big\rangle&=-\frac{\Omega^2(x_0^2+\ell^2m_2)}{2},
  \label{eq:magtrial}
\end{align}
so the internal terms $\pm\half\Omega^2\ell^2m_2$ cancel.

\emph{Expansions.} For $V$ smooth near $\x_0$, isotropy of $Q^2$ gives
\begin{equation}
  \avg{V(|\x_0+\ell\y|)}_Q=V(x_0)+\frac{\ell^2m_2}{4}\nabla^2V(x_0)+O(\ell^4),
  \label{eq:Vexp}
\end{equation}
because odd moments vanish and $\avg{y_iy_j}_Q=\half m_2\delta_{ij}$. With $\x_0$ on the
ring of $\Veff$ this gives \cref{eq:ring}; with $\x_0$ small and $\ell$ fixed, the
quadratic term in $\x_0$ gives \cref{eq:wvar}. For $x_0=0$ the average is
$m_s\ell^s/s$, which gives \cref{eq:Eell}; with the regularized core,
$\avg{V_{s,\eps}(\ell|\y|)}_Q=\eps^s/s+m_1\eps^{s-1}\ell+O(\ell^2)$, which gives
\cref{eq:cone}. In the computations all averages are evaluated by quadrature with the
actual trap, and \cref{eq:Eoff} is minimized over $(\ell,x_0)$ without expansion.

\emph{Relation to Guo, Li, and Luo.} Dividing the functional of Ref.~\cite{GuoLiLuo2024}
by two maps it onto \cref{eq:E} with the same $\Omega$, $g=-a/2$ (their $a$), and the trap
$V=\Omega^2(r^4/16+3r^2/8+1/16)$, whose $\Veff$ vanishes on the ring $r_0=1$ and for which
$\nabla^2V(r_0)=5\Omega^2/2$. The first form of \cref{eq:ring} then gives
$\ell_\ast^4=4\delta/(5m_2\Omega^2)$. This is the fourth power of the concentration scale of
their Theorem~1.2 at fixed rotation, $(g_{*}^{\rm GLL}-a)/(\Omega^2\lambda^4)$, with
$\lambda^4=\tfrac54\int_{\R^2}|\x|^2Q(\x)^2\,\dd\x$, $g_{*}^{\rm GLL}=\|Q\|_2^2$, and
$\delta=1-a/g_{*}^{\rm GLL}$.

\section{Thomas--Fermi scales}
\label{app:tf}
For $s>0$ the kinetic-free density $n^{\rm TF}=(\mu^{\rm TF}_0-r^s/s)_+/g$ is supported on
$r\le R^{\rm TF}_0=(s\mu^{\rm TF}_0)^{1/s}$, and its normalization,
\begin{equation}
  2\pi\int_0^{R^{\rm TF}_0}n^{\rm TF}(r)\,r\,\dd r=\frac{\pi\,(R^{\rm TF}_0)^{s+2}}{g(s+2)}=1,
\end{equation}
yields \cref{eq:RTF}. For the rotating quartic trap at $\mu^{\rm TF}=0$ the density
$(\Omega^2r^2/2-r^4/4)_+/g$ is supported on $r\le\sqrt2\,\Omega$, and
\begin{equation}
  \frac{2\pi}{g}\int_0^{\sqrt2\Omega_c}\Big(\frac{\Omega_c^2r^2}{2}-\frac{r^4}{4}\Big)r\,\dd r
  =\frac{\pi\Omega_c^6}{3g}=1
\end{equation}
gives $\Omega_c=(3g/\pi)^{1/6}$.

\section{Numerical methods and checks}
\label{app:numerics}

\emph{Two-dimensional solver.} All two-dimensional states are computed with one unmodified
version of the solver, archived with the data. The base step is $\Delta\tau=5\times10^{-4}$,
the energy is checked every $1000$ steps, and a state is energy stationary when
$|\Delta E/E|/(1000\,\Delta\tau)<2\times10^{-5}$ after a minimum imaginary time: $5$ for the
attractive scans, and $10$ for the first regulator of an endpoint series and $3$ for the
later ones, each of which starts from the state at the previous regulator. If $\etaGP$
exceeds its gate, the state is continued at the smaller steps
$\Delta\tau=2,1,0.5\times10^{-4}$ (scaled with the initial step
$\min(5\times10^{-4},0.2\eps^2)$ for the endpoint). On graphics processors the bulk relaxation runs in single precision;
every state is then completed in double precision, in which all reported observables are
evaluated. Attractive runs start from the variational Townes trial, with $10^{-3}$ azimuthal
noise for centered states; endpoint states are interpolated spectrally to the next
regulator.

\emph{Observables.} The width uses the internal kinetic energy,
\begin{equation}
  E^{\rm int}_{\rm kin}=E_{\rm kin}-\half|\avg{\mathbf p}|^2,\qquad
  \avg{\mathbf p}=\frac{\sum_{\mathbf k}\mathbf k\,|\hat\psi_{\mathbf k}|^2}{\sum_{\mathbf k}|\hat\psi_{\mathbf k}|^2},
  \label{eq:Ekinint}
\end{equation}
where $E_{\rm kin}=\half\|\nabla\psi\|_2^2$, $\hat\psi_{\mathbf k}$ are the discrete Fourier
coefficients, and $\mathbf k$ runs over the reciprocal grid $(2\pi/\Lsq)(n_x,n_y)$ with
integers $n_x,n_y$. The center of mass is $\avg{\x}$, with
$r_{\rm cm}=|\avg{\x}|$, and a state is flagged as concentrated to the grid when
$\ell<2.5\Delta x$. The central winding is the plaquette circulation inside $r<0.5$. In
the repulsive singular traps the net winding is
\begin{equation}
  Q_\phi=\mathrm{round}\Big[\frac1{2\pi}\Big|\sum_{p:\,r_p<0.96\Rbox}\Gamma_p\Big|\Big],
  \label{eq:Qphi}
\end{equation}
with $\Gamma_p$ the phase circulation around plaquette $p$ centered at $r_p$; $\Ncore$
counts unit-winding plaquettes whose corner density exceeds $10^{-3}$ of the peak. The
GP and TF densities are normalized independently, averaged azimuthally ($\bar n$) in
$\max(200,N_{\rm g}/2)$ radial bins, and compared through
\begin{equation}
  \delta n_{2,r}=\Bigg[\frac{\int_0^{\Rbox}|\bar n^{\rm GP}-\bar n^{\rm TF}|^2\,r\,\dd r}
  {\int_0^{\Rbox}|\bar n^{\rm TF}|^2\,r\,\dd r}\Bigg]^{1/2}.
  \label{eq:dn}
\end{equation}

\emph{Radial solver.} The radial energy
\begin{equation}
  E_{\rm rad}[f]=2\pi\int_0^{R_{\rm end}}\Big[\half f'^2+(V+V^{\rm wall})f^2+\frac g2f^4\Big]\,r\,\dd r,
  \label{eq:Erad}
\end{equation}
minimized with $2\pi\int_0^{R_{\rm end}}f^2r\,\dd r=1$ and $f(R_{\rm end})=0$, $R_{\rm end}=\Lsq/2$, uses
exactly the potentials of the two-dimensional runs (for the endpoint rescaled problem,
$V=W$ and no wall). The grid is uniform with spacing $h_0$ up to the smaller of $\eps$ and
the expected width and geometric beyond; potentials are averaged over each cell by
six-point Gauss--Legendre quadrature with weight $2\pi r$, which resolves the conical core
of $(r+\eps)^s$ and the onset of the wall. A backward-Euler normalized gradient flow is followed by a bordered Newton
iteration on $(H-\mu)f=0$, $\|f\|_2=1$, to a relative residual $\le10^{-8}$. Every value is
computed on three nested grids and Richardson extrapolated (observed order $2.00$).
Points whose relative energy uncertainty exceeds $3\times10^{-3}$ are discarded; this
uncertainty grows as $\delta^{-1}$ because kinetic and contact energies cancel to order
$\delta$. \Cref{tab:bench} compares the two-dimensional states with these references.

\begin{table*}[t]
\caption{Two-dimensional vs radial states ($\Omega=0$): number of compared states and
maximum relative differences. For the endpoint the energy column excludes points with
$|E\eps^2|<5\times10^{-3}$; there $|\Delta(E\eps^2)|\le1.3\times10^{-4}$.}
\label{tab:bench}
\begin{ruledtabular}
\begin{tabular}{lccccc}
Program, $s$ & States & $|\Delta E/E|$ & $|\Delta\ell/\ell|$ & $|\Delta r_{\rm rms}/r_{\rm rms}|$ & $|\Delta\Mcore/\Mcore|$\\ \hline
endpoint, $-1.5$ & 24 & $8.5\times10^{-4}$ & $3.3\times10^{-4}$ & $2.2\times10^{-4}$ & $1.1\times10^{-2}$ \\
endpoint, $-1.9$ & 24 & $2.5\times10^{-3}$ & $1.2\times10^{-3}$ & $9.7\times10^{-4}$ & $1.1\times10^{-2}$ \\
endpoint, $-2$ & 30 & $3.1\times10^{-3}$ & $1.6\times10^{-3}$ & $1.3\times10^{-3}$ & $1.1\times10^{-2}$ \\
concentration, $4$ & 8 & $1.9\times10^{-3}$ & $2.3\times10^{-2}$ & $2.2\times10^{-2}$ & -- \\
concentration, $3$ & 8 & $2.5\times10^{-3}$ & $3.0\times10^{-2}$ & $2.9\times10^{-2}$ & -- \\
concentration, $2$ & 8 & $1.0\times10^{-3}$ & $2.3\times10^{-2}$ & $2.2\times10^{-2}$ & -- \\
concentration, $1$ & 8 & $1.1\times10^{-3}$ & $3.5\times10^{-2}$ & $3.5\times10^{-2}$ & -- \\
concentration, $-0.5$ & 18 & $2.3\times10^{-4}$ & $8.6\times10^{-3}$ & $8.5\times10^{-3}$ & -- \\
concentration, $-1$ & 18 & $2.1\times10^{-3}$ & $3.0\times10^{-2}$ & $3.0\times10^{-2}$ & -- \\
concentration, $-1.5$ & 17 & $1.1\times10^{-2}$ & $4.5\times10^{-2}$ & $4.4\times10^{-2}$ & -- \\
\end{tabular}
\end{ruledtabular}
\end{table*}

\subsection{Bogoliubov frequencies of centered states}
\label{app:bogo}
On the cell-centered grid $r_j=(j-\half)h$, $j=1,\dots,n$, $nh=R$ ($R=5$ for $s=3,4$),
the quadratic forms $\half\int_0^R(f'^2+m^2f^2/r^2)\,2\pi r\,\dd r$ are discretized by flux
differences with zero flux at $r=0$ and $f(R)=0$; $\psi_0$ and $\mu$ are obtained on the
same grid by gradient flow and Newton iteration. With the weight $w_j=2\pi r_jh$ and the
symmetric matrices $K_\pm$ of the discretized quadratic forms
$\int_0^R f\,(L_\pm^{(m)}f)\,2\pi r\,\dd r$, the matrices $A_\pm=w^{-1/2}K_\pm w^{-1/2}$
represent $L_\pm^{(m)}$, and the
frequencies are the positive eigenvalues $\omega$ of the pencil
\begin{equation}
  \begin{pmatrix}A_+&0\\0&A_-\end{pmatrix}\mathbf v=\omega\begin{pmatrix}0&1\\1&0\end{pmatrix}\mathbf v ,
  \label{eq:pencil}
\end{equation}
equivalent to $A_-A_+\mathbf v_1=\omega^2\mathbf v_1$ for the upper block $\mathbf v_1$ of $\mathbf v$, and
found by shift-invert iteration with a sparse LU factorization. Forming the product
$A_-A_+$ instead squares the condition number and loses accuracy below $h=10^{-3}$. The
lowest $\omega_m$ converge at second order (successive-difference ratio $4.00$) and are
Richardson extrapolated over $h=10^{-3},5\times10^{-4},2.5\times10^{-4}$ for $\delta<0.05$ and
twice those spacings otherwise; the remaining uncertainty is below $6\times10^{-5}$. The
radial ground states reproduce the reference widths and energies (e.g., $\ell=0.58373$,
$E=0.238231$ at $\delta=0.1$). For $m=1$ the harmonic trap gives the Kohn value to
$3\times10^{-7}$. For $s=4$ and $\delta=0.003$--$1$, $\omega_2/2\ge1.079\,\omega_1$ and
$\omega_3/3\ge1.147\,\omega_1$; for $s=3$ over the same range, $\omega_2/2\ge1.049\,\omega_1$ and
$\omega_3/3\ge1.089\,\omega_1$, the minima occurring at $\delta=1$ in both cases. The threshold
$\dGP(\Omega)$ is located by Brent's method on $\omega_1(\delta)-\Omega$; recomputing
$\omega_1$ at the tabulated $\dGP$ for $s=3$ returns $\Omega$ to $3\times10^{-5}$.

\subsection{Stationary bias of the split step}
\label{app:dtau}
Write one step of \cref{eq:split} at $\Omega=0$ as
$\psi_1=e^{-\Delta\tau U/2}\psi$, $\psi_2=e^{-\Delta\tau T}\psi_1$, $\psi_3=e^{-\Delta\tau U'/2}\psi_2$,
$T=-\half\nabla^2$, followed by normalization, with $U=V+g|\psi|^2$ and $U'=V+g|\psi_2|^2$. At a
fixed point $(T+U)\psi=\mu\psi+O(\Delta\tau)$, so
$\psi_2=\psi-\Delta\tau(\mu-\half U)\psi+O(\Delta\tau^2)$ and
$U'=U-\Delta\tau\,\delta U+O(\Delta\tau^2)$ with $\delta U=g(2\mu-U)|\psi|^2$. By the symmetric
Baker--Campbell--Hausdorff formula the step operator is then
\begin{align}
  &e^{\Delta\tau^2\delta U/2}\,e^{-\Delta\tau U/2}e^{-\Delta\tau T}e^{-\Delta\tau U/2}\nonumber\\
  &\quad=\exp\!\big[-\Delta\tau\big(T+U-\tfrac12\Delta\tau\,\delta U\big)+O(\Delta\tau^3)\big].
  \label{eq:BCH}
\end{align}
The fixed point is
therefore the ground state of \cref{eq:Hdt}. If $U'$ is computed after renormalizing
$\psi_2$, $\delta U$ is shifted by its mean and remains nonuniform, so the bias stays first
order; with $U'=U$ the product is the symmetric Strang product for a frozen potential,
whose stationary state differs from the ground state at $O(\Delta\tau^2)$. The prediction in
\cref{tab:dtau} solves \cref{eq:Hdt} self-consistently with the symmetric step at
$\Delta\tau=2.5\times10^{-4}$. With rotation, $P_2$ approximates $e^{\Omega\Delta\tau\Lz}$ to
$O(\Delta\tau^3)$ per step and so adds a term of order $\Delta\tau^2$ to the stationary
bias.

\subsection{Rotating quartic runs}
\label{app:ring}
Each point of \cref{fig:rot} was recomputed on the production grid at
$\Delta\tau=8,4,2,1\times10^{-4}$, each stage starting from the previous one, and ran until the width was
stationary by a geometric-tail estimate or until a time limit ($\tau=80$ for the first stage
and $40$ otherwise). The time limit was reached in all stages of the displaced states at
$\Omega=1$ and in the last three stages of the centered states with $\delta\le0.0125$; all
other stages converged. Widths, centers, and angular momenta are extrapolated as
$X_0+a\Delta\tau+b\Delta\tau^2$ and energies as $E_0+b\Delta\tau^2+c\Delta\tau^3$ from the
three smallest steps. \Cref{tab:ring} lists the extrapolated states and the largest estimated
remaining relative drift of the width over the stages, $d_\ell$. For the time-limited
displaced states at $\Omega=1$ this drift is at most $3.4\times10^{-5}$, except at
$\delta=0.3$, where the slow radial relaxation of the center near the transition prevents an
estimate; that state is used only together with the Bogoliubov result of \cref{sec:rot}.
As an uncertainty we take the difference between the extrapolations from the three largest
and the three smallest steps. Relative to the extrapolated values it is below
$2.2\times10^{-3}$ in $\ell$ for every listed state except the harmonic control at
$\delta=0.00625$ ($4.1\times10^{-3}$), below $1.5\times10^{-4}$ for $\delta\ge0.025$, and below
$7\times10^{-4}$ in $E$.

For a grid check we use a CPU replica of the step \cref{eq:split}. Started from the archived
$\Omega=2$, $\delta=0.05$ state, it reproduces the production stage at
$\Delta\tau=4\times10^{-4}$ to $3\times10^{-6}$ in $\ell$. On a square of side $8$ with $640^2$
points ($\Delta x=0.0125$ instead of $0.0234$) the same run differs from that on the
production grid by less than $3\times10^{-8}$ (relative) in $\ell$, $r_{\rm cm}$, $E$, and
$\avg{\Lz}$ at equal imaginary time, a difference that is still decreasing as the
interpolation transient decays.

\begin{table*}[t]
\caption{Rotating quartic trap ($s=4$) and nonrotating controls ($\Omega=0$, including one
harmonic control, $s=2$), extrapolated to $\Delta\tau\to0$: width $\ell$ and its ratio to the
ring law \cref{eq:ring} (displaced states) or to \cref{eq:conc} (centered states), radial
reference width $\ell_{\rm rad}$, center $r_{\rm cm}$ and trial center $x_{0,\rm var}$ in units
of $r_0=\Omega$, energy $E$ and trial bound $\Evar$, and estimated remaining drift $d_\ell$.
At $\Omega=1$, $\delta=0.3$ both the centered and the displaced stationary states are
listed.}
\label{tab:ring}
\begin{ruledtabular}
\begin{tabular}{ccccccccccc}
$s$ & $\Omega$ & $\delta$ & $\ell$ & $\ell/\ell_\ast$ & $\ell_{\rm rad}$ & $r_{\rm cm}/r_0$ & $x_{0,\rm var}/r_0$ & $E$ & $\Evar$ & $d_\ell$\\ \hline
2 & 0 & 0.00625 & 0.27049 & 1.004 & 0.27013 & --- & --- & 0.08599 & 0.08615 & $8\times10^{-8}$ \\
4 & 0 & 0.00625 & 0.34170 & 1.017 & 0.34162 & --- & --- & 0.04090 & 0.04151 & $2\times10^{-7}$ \\
4 & 0 & 0.0125 & 0.38766 & 1.028 & 0.38766 & --- & --- & 0.06423 & 0.06589 & $1\times10^{-9}$ \\
4 & 0 & 0.025 & 0.44208 & 1.044 & 0.44208 & --- & --- & 0.10027 & 0.10459 & $1\times10^{-9}$ \\
4 & 0.5 & 0.0125 & 0.38766 & 1.028 & 0.38766 & 0 & 0 & 0.06423 & 0.06589 & $1\times10^{-9}$ \\
4 & 0.5 & 0.025 & 0.44208 & 1.044 & 0.44208 & 0 & 0 & 0.10027 & 0.10459 & $1\times10^{-9}$ \\
4 & 1 & 0.3 & 0.72913 & 1.138 & --- & 0 & 0.2096 & 0.46031 & 0.54795 & $6\times10^{-9}$ \\
4 & 1 & 0.3 & 0.72254 & 1.212 & --- & 0.1775 & 0.2096 & 0.46018 & 0.54795 & --- \\
4 & 1 & 0.2 & 0.62722 & 1.164 & --- & 0.4841 & 0.4911 & 0.35345 & 0.40920 & $4\times10^{-6}$ \\
4 & 1 & 0.1 & 0.50056 & 1.105 & --- & 0.6909 & 0.6932 & 0.19858 & 0.22278 & $9\times10^{-7}$ \\
4 & 1 & 0.05 & 0.40549 & 1.065 & --- & 0.7994 & 0.8001 & 0.07772 & 0.08746 & $2\times10^{-6}$ \\
4 & 1 & 0.025 & 0.33260 & 1.038 & --- & 0.8653 & 0.8654 & $-0.01352$ & $-0.00986$ & $8\times10^{-6}$ \\
4 & 1 & 0.0125 & 0.27545 & 1.023 & --- & 0.9078 & 0.9078 & $-0.08075$ & $-0.07945$ & $3\times10^{-5}$ \\
4 & 2 & 0.3 & 0.45700 & 1.084 & --- & 0.9475 & 0.9451 & $-2.44515$ & $-2.32220$ & $6\times10^{-9}$ \\
4 & 2 & 0.2 & 0.40744 & 1.070 & --- & 0.9571 & 0.9555 & $-2.70460$ & $-2.62851$ & $4\times10^{-9}$ \\
4 & 2 & 0.1 & 0.33509 & 1.046 & --- & 0.9696 & 0.9688 & $-3.06116$ & $-3.02877$ & $6\times10^{-9}$ \\
4 & 2 & 0.05 & 0.27699 & 1.028 & --- & 0.9784 & 0.9781 & $-3.32563$ & $-3.31252$ & $8\times10^{-9}$ \\
4 & 2 & 0.025 & 0.23024 & 1.017 & --- & 0.9847 & 0.9846 & $-3.51858$ & $-3.51352$ & $2\times10^{-8}$ \\
4 & 2 & 0.0125 & 0.19233 & 1.010 & --- & 0.9892 & 0.9891 & $-3.65770$ & $-3.65583$ & $4\times10^{-7}$ \\

\end{tabular}
\end{ruledtabular}
\end{table*}

\section{Endpoint checks}
\label{app:endpoint}
The rescaled problem \cref{eq:resc} is solved with the radial solver on $D_\Lambda$ with a
Dirichlet condition, on three nested grids with a uniform core of $60$--$240$ cells per
unit $u$ and geometric growth to $\Lambda$. Results for $\Lambda=10^6$ and $10^7$ ($10^8$ and $10^{10}$
for $g\ge30$) agree within their uncertainties, which are estimated from the three grids
and stay below $4\times10^{-6}$ (relative) for $e$, $\Mcore$, and $L$. For $g\ge40$ the chemical
potential is $|\mu|\lesssim3\times10^{-8}$, and the relative residual \cref{eq:res}, normalized
by $|\mu|$, saturates at round-off, $3\times10^{-5}$--$6\times10^{-4}$, without affecting these
uncertainties. \Cref{tab:wall} compares the soft wall of
the computations with a hundredfold stiffer wall and with the Dirichlet disk of
Proposition~1 at $s=-2$, $\Rbox=2$. \Cref{tab:kappa} gives the rescaled problem for the tail
strengths $\kappa=1/2$ and $2$ with the same solver and $\Lambda=10^7$; the bounds $-\kappa/2$ and
$-\pi\kappa^2/(24g)$ hold in every case.

\begin{table}[t]
\caption{Rescaled endpoint problem for the tail $-\kappa/(2u^2)$, $W\to\kappa W_1$: energy,
core mass ($u<2$), rms size, $\ln C=\ln L_h-g/[\pi(1+\kappa)]$ with
$L_h=[2g/\pi(1+\kappa)]^{1/2}L$, and the maximum of $g\,u^2|\phi_g|^2$, whose plateau is
predicted at $(1+\kappa)/2$ [\cref{eq:halokappa}].}
\label{tab:kappa}
\setlength{\tabcolsep}{3pt}
\begin{ruledtabular}
\begin{tabular}{cccccc}
$\kappa$, $g$ & $-e$ & $\Mcore$ & $L$ & $\ln C$ & $\max g\,u^2|\phi_g|^2$\\ \hline
1/2, 5 & 0.001017 & 0.02266 & 38.48 & 2.965 & 0.327\\
1/2, 10 & 0.0005509 & 0.01222 & 102.7 & 3.233 & 0.521\\
1/2, 20 & 0.0002779 & 0.006160 & 684.3 & 3.354 & 0.688\\
1/2, 30 & 0.0001853 & 0.004107 & 4737 & 3.369 & 0.734\\
2, 5 & 0.04045 & 0.2482 & 5.526 & 1.209 & 0.403\\
2, 10 & 0.02550 & 0.1535 & 9.368 & 1.552 & 0.711\\
2, 20 & 0.01380 & 0.08210 & 24.40 & 1.795 & 1.103\\
2, 30 & 0.009278 & 0.05509 & 61.95 & 1.869 & 1.304\\
2, 50 & 0.005573 & 0.03308 & 413.8 & 1.901 & 1.453\\
2, 80 & 0.003483 & 0.02068 & 7924 & 1.905 & 1.494\\
\end{tabular}
\end{ruledtabular}
\end{table}

\begin{table}[t]
\caption{Wall dependence at $s=-2$, $\Rbox=2$: soft wall $\kappa_{\rm w}=5000$ (used in the
computations), $\kappa_{\rm w}=5\times10^5$, and Dirichlet disk. The first two cases fit in the
box; the third is box limited in its halo only, the last two in core and halo.}
\label{tab:wall}
\setlength{\tabcolsep}{4pt}
\begin{ruledtabular}
\begin{tabular}{ccccc}
$g$, $\eps$ & wall & $E\eps^2$ & $\Mcore$ & $r_{\rm rms}/\eps$\\ \hline
2, 0.025 & $5\times10^3$ & $-1.1949\times10^{-2}$ & 0.14529 & 8.00 \\
 & $5\times10^5$ & $-1.1949\times10^{-2}$ & 0.14529 & 8.00 \\
 & Dirichlet & $-1.1949\times10^{-2}$ & 0.14529 & 8.00 \\
10, 0.01 & $5\times10^3$ & $-4.3488\times10^{-3}$ & 0.05098 & 27.06 \\
 & $5\times10^5$ & $-4.3488\times10^{-3}$ & 0.05098 & 27.04 \\
 & Dirichlet & $-4.3488\times10^{-3}$ & 0.05098 & 27.03 \\
50, 0.002 & $5\times10^3$ & $-8.9294\times10^{-4}$ & 0.01047 & 443.46 \\
 & $5\times10^5$ & $-8.9253\times10^{-4}$ & 0.01047 & 425.01 \\
 & Dirichlet & $-8.9231\times10^{-4}$ & 0.01047 & 416.52 \\
10, 0.04 & $5\times10^3$ & $-4.1660\times10^{-3}$ & 0.05302 & 18.35 \\
 & $5\times10^5$ & $-4.1265\times10^{-3}$ & 0.05335 & 17.82 \\
 & Dirichlet & $-4.1051\times10^{-3}$ & 0.05352 & 17.57 \\
50, 0.02 & $5\times10^3$ & $-1.6493\times10^{-4}$ & 0.01142 & 55.69 \\
 & $5\times10^5$ & $-8.0141\times10^{-5}$ & 0.01151 & 53.21 \\
 & Dirichlet & $-3.6418\times10^{-5}$ & 0.01156 & 52.07 \\

\end{tabular}
\end{ruledtabular}
\end{table}

\section{Repulsive diagnostics}
\label{app:rep}
\Cref{tab:epsgrad} lists the archived density-gradient ratios $\epsgrad$
[\cref{eq:epsgrad}] in harmonic and quartic traps (\cref{sec:rot-rep}); the flow part was
not recorded for these states. The harmonic benchmark of \cref{sec:rot-rep} was recomputed
with a CPU replica of the step \cref{eq:split} on $256^2$ points on a square of side $20$
($\Delta x=0.078$, about $0.4$ healing lengths at the center), started from an imprinted
lattice of $15$ vortices and relaxed at $\Delta\tau=10^{-3}$ to $\tau=200$
($\etaGP=2.0\times10^{-3}$). Six vortices enter from the edge during the first $\tau\approx15$;
at the end the energy changes by less than $10^{-6}$ per unit $\tau$ (relative). Whether a
lattice with another number of vortices lies lower was not tested. Between $\tau=80$ and
$200$, $\epsgrad$ and $\gflow$ change by less than $10^{-3}$ and the kinetic ratio by
$0.007$. The density-gradient ratio of the $21$-vortex
lattice, $0.057$, is close to the archived value $0.055$ of a $15$-vortex lattice at the same
parameters (kinetic ratio $0.917$), for which $\gflow$ is not available.

The design of \cref{fig:reprot}(a) has $30$ energy-stationary states at $\Omega=0.3$
($s=-0.25$ to $-1.99$, $g=500$, $1000$, $2000$); $\etaGP$ grows toward $s=-2$. \Cref{tab:cal}
calibrates the production grid at $\Omega=0$, where the ground state is radial: the same
step and potential on the production grid (and on $1024^2$ points for some cases) are
compared with radial solutions on graded grids refined toward the cusp and Richardson
extrapolated over three grids (relative uncertainties below $3\times10^{-6}$ for $\epsgrad$ and
$8\times10^{-4}$ for $\delta n_{2,r}$). The replica reproduces the archived production
states: at $s=-1.9$, $g=500$ the $\Omega=0.3$ state carries no vortex
($\avg{\Lz}=7\times10^{-6}$) and coincides with the $\Omega=0$ state, and the replica gives
its $\epsgrad$ and $\etaGP$ to six digits. At $\Delta\tau=5\times10^{-4}$ the residual near
the cusp is dominated by the stationary bias of the step (\cref{sec:dtau-num}): at
$s=-1.75$, $g=1000$, quartering $\Delta\tau$ lowers it from $0.15$ to $0.014$ on the
production grid and from $0.38$ to $0.041$ on $1024^2$ points, while the errors of
$\epsgrad$ and $\delta n_{2,r}$ change by at most $2$ percentage points ($14\%$ and $45\%$ on
$512^2$, $1.5\%$ and $15\%$ on $1024^2$ points). At fixed $\Delta\tau$ the residual therefore
grows with the resolved depth of the cusp, and it measures the grid error only on the
production grid, where both grow toward $s=-2$. \Cref{tab:corr} gives
the rank correlations of $\epsgrad$ with five TF-error measures: $\delta n_{2,r}$, the $L^1$
distances of the azimuthally averaged and of the two-dimensional densities, and the relative
differences of the rms radius and of the radius that encloses $90\%$ of the atoms. The
controlled coefficient is the partial rank correlation with the rank of $g$ removed.

The branch check of \cref{sec:rot-rep} uses the replica with rotation at $s=-1.5$, $g=2000$,
$\Omega=0.3$, started from the rotating-frame TF density with seven imprinted vortices (one at
the center and a hexagon of radius $3.5$) and relaxed at the production step to $\tau=20$,
where $\etaGP=0.015$, as for the archived state; the winding stays at $Q_\phi=7$. The
rotating analysis restricts the TF density to $r\le\Rbox$, which reproduces the archived
$\mu^{\rm TF}$; the calibration of \cref{tab:cal} includes the soft-wall region, and the two
definitions differ by about $1\%$ in $\delta n_{2,r}$. Continued to $\tau=22$ on $1024^2$
points, the same state changes $\epsgrad$ by $-0.5\%$, $\gflow$ by less than $10^{-4}$
(relative), and $\delta n_{2,r}$, evaluated in the same 256 bins, by $-6\%$, the size of the
grid errors found at $\Omega=0$ for the same $s$ and $g$.

\begin{table}[t]
\caption{Calibration at $\Omega=0$ of the production step ($\Delta\tau=5\times10^{-4}$) for
the boxed repulsive singular traps ($\eps=0.02$, $\Rbox=7$) on $N_{\rm g}^2$ points: residual
of the two-dimensional state and relative errors, in percent, of $\epsgrad$ and of the binned
$\delta n_{2,r}$ against converged radial solutions. $N_{\rm g}=512$ ($\eps/\Delta x=0.51$) is
the production grid; for $N_{\rm g}=1024$ the annular average uses 512 bins, and the radial
reference is binned the same way.}
\label{tab:cal}
\setlength{\tabcolsep}{4pt}
\begin{ruledtabular}
\begin{tabular}{rrrccc}
$s$ & $g$ & $N_{\rm g}$ & $\etaGP$ & $\Delta\epsgrad$ (\%) & $\Delta\delta n_{2,r}$ (\%)\\ \hline
$-1$ & 500 & 512 & 0.0085 & $+0.066$ & $+1.7$\\
$-1.75$ & 500 & 512 & 0.37 & $+35$ & $+54$\\
$-1.9$ & 500 & 512 & 0.78 & $+106$ & $+48$\\
$-0.5$ & 1000 & 512 & 0.00084 & $-0.028$ & $+0.18$\\
$-1$ & 1000 & 512 & 0.0037 & $+0.0041$ & $+0.85$\\
$-1$ & 1000 & 1024 & 0.012 & $-0.026$ & $+0.18$\\
$-1.25$ & 1000 & 512 & 0.010 & $+0.19$ & $+3.6$\\
$-1.5$ & 1000 & 512 & 0.034 & $+1.7$ & $+15$\\
$-1.5$ & 1000 & 1024 & 0.11 & $+0.23$ & $+3.8$\\
$-1.75$ & 1000 & 512 & 0.15 & $+16$ & $+44$\\
$-1.75$ & 1000 & 1024 & 0.38 & $+2.3$ & $+14$\\
$-1.9$ & 1000 & 512 & 0.43 & $+61$ & $+54$\\
$-1.9$ & 1000 & 1024 & 0.72 & $+9.8$ & $+25$\\
$-1.95$ & 1000 & 512 & 0.58 & $+88$ & $+52$\\
$-1$ & 2000 & 512 & 0.0017 & $-0.037$ & $+0.66$\\
$-1.25$ & 2000 & 512 & 0.0046 & $+0.047$ & $+2.1$\\
$-1.5$ & 2000 & 512 & 0.015 & $+0.67$ & $+8.6$\\
$-1.75$ & 2000 & 512 & 0.061 & $+6.8$ & $+30$\\
$-1.9$ & 2000 & 512 & 0.18 & $+27$ & $+45$\\
\end{tabular}
\end{ruledtabular}
\end{table}

\begin{table}[t]
\caption{Spearman rank correlation $\rho$ of $\epsgrad$ with TF-error measures, and the
partial coefficient $\rho_g$ with the rank of $g$ controlled, for the 14 gated states at
$\Omega=0.3$ ($\etaGP\le0.02$) and for 30 converged radial states at $\Omega=0$
($s=-0.25$ to $-1.99$, $g=500$, $1000$, $2000$; the two $L^1$ measures coincide there).}
\label{tab:corr}
\begin{ruledtabular}
\begin{tabular}{lcccc}
 & \multicolumn{2}{c}{$\Omega=0.3$, gated} & \multicolumn{2}{c}{$\Omega=0$, radial}\\
Measure & $\rho$ & $\rho_g$ & $\rho$ & $\rho_g$\\ \hline
$\delta n_{2,r}$ & 0.952 & 0.627 & 0.990 & 0.966\\
$L^1$, azimuthal average & 0.864 & $-0.051$ & 1.000 & 1.000\\
$L^1$, two-dimensional & 0.868 & $-0.017$ & 1.000 & 1.000\\
rms radius & 0.749 & $-0.932$ & 0.992 & 0.931\\
$90\%$ radius & 0.881 & $-0.504$ & 0.831 & $-0.535$\\
\end{tabular}
\end{ruledtabular}
\end{table}

\begin{table}[t]
\caption{Density-gradient ratio $\epsgrad$ of rotating repulsive states (archived values;
the flow part $\gflow$ was not recorded for these states).}
\label{tab:epsgrad}
\begin{ruledtabular}
\begin{tabular}{cccc}
\multicolumn{4}{c}{$s=2$}\\
$\Omega$ & $g=500$ & $g=1500$ & $g=3000$\\ \hline
0.3 & 0.028 & 0.018 & 0.010\\
0.5 & 0.055 & 0.027 & 0.018\\
0.7 & 0.085 & 0.047 & 0.031\\
0.9 & 0.192 & 0.097 & 0.067\\ \hline
\multicolumn{4}{c}{$s=4$}\\
$\Omega$ & $g=500$ & $g=1000$ & $g=2000$\\ \hline
0.5 & 0.034 & 0.018 & 0.010\\
0.9 & 0.034 & 0.031 & 0.017\\
1.5 & 0.068 & 0.044 & 0.023\\
2.0 & 0.102 & 0.062 & 0.031\\
\end{tabular}
\end{ruledtabular}
\end{table}

\section{Machine-checked statements}
\label{app:lean}
The algebraic and calculus steps of \cref{sec:theory} were checked with the proof assistant
Lean~4 and its mathematical library, using only the standard axioms of its logic: the trial
energies \cref{eq:Eell,eq:quartic,eq:Eoff} and their minima, including
\cref{eq:conc,eq:cone,eq:dcvar,eq:ring}, the continuity of the transition at $\dvar$, the
Guo--Li--Luo mapping, the $\eps^{-2}$ covariance and the Young bounds of Proposition~1, the
halo solution and normalization \cref{eq:halo,eq:Lh} for $\kappa=1$ (the extension to general
$\kappa$, \cref{eq:halokappa}, is not machine checked), the TF normalizations, and the
moment $\avg{r^2}=(m+1)/\Omega$ behind \cref{eq:LLL}. The Gagliardo--Nirenberg inequality,
Jensen's inequality, and the Pohozaev identities enter as hypotheses (numerically $\|\nabla Q\|_2^2/\|Q\|_2^2=1-6\times10^{-11}$ and
$\|Q\|_4^4/2\|Q\|_2^2=1+3\times10^{-9}$).

\section*{Data availability}
The two-dimensional solver, the radial and Bogoliubov solvers, the tables of every computed
state (including the full scans summarized in \cref{tab:bench}), the figure scripts, the
formal proofs, and the scripts and outputs of the CPU checks (grid refinement of the ring
states, radial references and grid calibration for the repulsive singular traps, the
harmonic vortex lattice, and the tail-strength computations) are provided in a reproducibility archive upon request.

\bibliographystyle{apsrev4-2}
\bibliography{PL}

\end{document}